\documentclass[11pt]{article}
\usepackage{latexsym,amssymb,amstext,amsmath}

\newcommand{\be}[1]{ \begin{equation}\label{#1} }
\newcommand{\ee}{\end{equation}}
\newcommand{\ben}[1]{\begin{eqnarray}\label{#1} }
\newcommand{\een}{\end{eqnarray}}

\newcommand{\refb}[1]{(\ref{#1})}
\newcommand{\OO}{{\cal O}}

\newcommand{\DD}{{\cal D}}

\newcommand{\LL}{{\cal L}}

\newcommand{\eb}{\bar{\epsilon}}

\newcommand{\ga}{\gamma}
\newcommand{\thc}{\text{h.c.}}
\newcommand{\Dslash}{\not{\hbox{\kern-4pt $D$}}}
\newcommand{\Pslash}{\not{\hbox{\kern-2.5pt $P$}}}
\newcommand{\pslash}{\not{\hbox{\kern-4pt $\partial$}}}
\newcommand{\Dcslash}{\not{\hbox{\kern-4pt $\DD$}}}

\usepackage{fancyhdr}
\usepackage{graphicx}
\usepackage{amsthm}
\usepackage{slashed}
\usepackage{color}
\usepackage[latin1]{inputenc} 
\usepackage[T1]{fontenc} 
\usepackage{multirow}
\usepackage[english]{babel}
\usepackage{eufrak}
\usepackage{bbold}
\usepackage{blkarray}
\usepackage{multirow}
\usepackage{hyperref}

\allowdisplaybreaks 
\def\Slash#1{\rlap{\hbox{$\mskip 3 mu /$}}#1}      

\begin{document}
%
\begin{titlepage}

\bigskip

\begin{center}
 {\LARGE\bfseries Towards the full $N=4$ conformal supergravity action} 
\\[10mm]
\textbf{Franz Ciceri$^a$, Bindusar Sahoo$^b$}\\[5mm] 
\vskip 4mm
$^a${\em Nikhef Theory Group, Science Park 105, 1098 XG Amsterdam, The
  Netherlands}\\
$^b${\em Indian Institute of Science Education and Research, Thiruvananthapuram}\\
{\em Kerala 695016, India} \\
  [3mm]
 {\tt f.ciceri@nikhef.nl}\,,\;{\tt bsahoo@iisertvm.ac.in}
 \end{center}

\vspace{3ex}

\begin{center}
{\bfseries Abstract}
\end{center}
\begin{quotation} \noindent
Based on the known non-linear transformation rules of the Weyl multiplet fields, the action of $N=4$ conformal supergravity is constructed up to terms quadratic in the fermion fields. The bosonic sector corrects a recent result in the literature.
\end{quotation}

\vfill

\end{titlepage}

\section{Introduction}
\label{sec:intro}
\setcounter{equation}{0}

Conformal supergravities in four dimensions are invariant under the local symmetries associated with the superconformal algebra $\,\mathfrak{su}(2,2\vert N)$. The transformation rules and corresponding invariant Lagrangians are known for $N=1$ and $2$ \cite{Kaku:1978nz,deWit:1979ug}. For the $N=4$ theory, the Weyl multiplet and its full non-linear transformations were determined in \cite{Bergshoeff:1980is}. A unique feature of the latter theory is the presence of scalars fields which parametrize an $\mathrm{SU}(1,1)/\mathrm{U}(1)$ coset space. This $\mathrm{U}(1)$ factor extends the $\mathrm{SU}(4)$ R-symmetry to the $\mathrm{U}(4)$ that is generically present in the algebra \cite{Ferrara:1977ij}. Furthermore, it was shown that $N>4$ theories cannot exist off-shell \cite{deWit:1978pd}, as they would necessarily involve higher-spin fields and the supermultiplet would in general not contain the graviton. It is also worth pointing out that the $N\leq 4$ superconformal field representation and the transformation rules have been worked out in superspace \cite{Howe:1981gz}. 

Although the field representation and its off-shell transformation rules are known, the full non-linear action for $N=4$ conformal supergravity remains to be constructed.  Recently, a calculation was performed based on an on-shell $N=4$ abelian gauge theory in a conformal supergravity background \cite{deRoo:1984gd}. The integration of the abelian gauge multiplet led to the determination of the bosonic terms of the superconformal action \cite{Buchbinder:2012uh}. These terms comprise the square of the Weyl tensor and are related to the conformal anomaly, as was discussed long ago in \cite{Fradkin:1981jc}. The resulting action is invariant under a continuous rigid $\mathrm{SU}(1,1)$ symmetry, which can be explained by the fact that the gauge theory action has $\mathrm{SU}(1,1)$ as an electric-magnetic duality group.  

In this paper we calculate the $\mathrm{SU}(1,1)$ invariant action of $N=4$ conformal supergravity by exploiting the known transformation rules and imposing supersymmetry by iteration.  This computation is of interest since it completes the result of \cite{Buchbinder:2012uh} to quadratic order in the fermion fields. However, we also find that our results do not coincide.

Actually, string theory indicates the existence of an extended class of actions in which the continuous $\mathrm{SU(1,1)}$ is broken. For instance, in IIA string compactifications on $\mathrm{K3}\times T^2$, the effective action contains terms quadratic in the Weyl tensor and its dual, multiplied by a modular function \cite{Harvey:1996ir}. Further indications arise from the semiclassical approximation of the microscopic degeneracy formula for dyonic BPS black holes \cite{LopesCardoso:2004xf,Jatkar:2005bh,LopesCardoso:2006bg}, which captures corrections to the macroscopic entropy originating from the same class of actions. This paper deals exclusively with the construction of the action invariant under the continuous $\mathrm{SU}(1,1)$.

The paper is organized as follows. Section \ref{sec:Preliminaries} contains a summary of the $N=4$ Weyl multiplet and its transformation rules. The quadratic action, which serves as the starting point of our computation, is discussed in section \ref{sec:quadratic-lagrangian}. In section \ref{sec:strategy}, we introduce the iterative procedure used to construct terms of higher-order in the fields. All the terms up to quadratic order in fermions are presented. Those that contain only matter fields, supercovariant derivatives and curvatures are discussed and compared to the result of \cite{Buchbinder:2012uh} in section \ref{sec:results}. The remaining terms which depend explicitly on the fermionic gauge fields are given in appendix \ref{App:Add-results}. Finally, appendix \ref{App:trans-cur} contains the Bianchi identities and the transformation rules of the curvatures.

\setcounter{equation}{0}

\section{$N=4$ conformal supergravity}
\label{sec:Preliminaries}
\setcounter{equation}{0}
$N=4$ conformal supergravity \cite{Bergshoeff:1980is} is built upon the gauging of the superconformal algebra $\mathfrak{su}(2,2\vert 4)$. Its bosonic subalgebra\footnote{The optional $\mathrm{U}(1)$ central charge is suppressed \cite{Ferrara:1977ij}. Note that it does not correspond to the one of the $\mathrm{SU}(1,1)/\mathrm{U}(1)$ coset space.} contains the generators of the conformal group $\mathrm{SU}(2,2)$ and the generators of a chiral $\mathrm{SU}(4)$ R-symmetry. The fermionic generators consist of sixteen Q supercharges and sixteen S supercharges. In addition, the theory has a non-linearly realised rigid $\mathrm{SU}(1,1)$ symmetry and a local chiral $\mathrm{U}(1)$ symmetry. The latter extends the R-symmetry group to $\mathrm{SU}(4)\times\mathrm{U}(1)$. The field representation of the theory comprises the gauge fields associated to the various superconformal symmetries and the local $\mathrm{U}(1)$, as well as a set of matter fields. In this paper, we adopt the conventions of \cite{Bergshoeff:1980is}, unless stated otherwise.

The bosonic gauge fields associated to the $\mathrm{SU}(2,2\vert 4)$ symmetries are the vierbein $e_\mu{}^a$, the spin connection $\omega_\mu{}^{ab}$, the dilatational gauge field $b_\mu$, the conformal boost gauge field $f_\mu{}^a$ and the $\mathrm{SU}(4)$ gauge field $V_\mu{}^i{}_j$, while the fermionic ones are the Q- and S-supersymmetry gauge fields $\psi_\mu{}^i$ and $\phi_\mu{}^i$, respectively. Finally, the connection $a_\mu$ is associated with the local chiral $\mathrm{U}(1)$ symmetry. The complete set of gauge fields of $N=4$ conformal supergravity is listed in table \ref{table:gaugefields} along with their algebraic restrictions, their $\mathrm{SU}(4)$ representation, their weight $w$ under local dilatations and their $\mathrm{U}(1)$ chiral weight $c$.\newline

\begin{table}[h!]
 \caption{
  Gauge fields of $N=4$ conformal supergravity}\label{table:gaugefields}
  \begin{center}
 \begin{tabular}{ c | c c l c c c }
 \hline 
 \hline
   & Field & Symmetries (Generators) & Name/Restrictions & $\mathrm{SU}(4)$ & $w$ & $c$ \\
 \hline
 \multirow{8}{*}{Bosons} & $e_\mu{}^a$ & Translations (P)& vierbein & 1 & $-1$ & 0 \\
 & $\omega_{\mu}{}^{ab}$  & Lorentz (M) & spin connection & 1 & 0 & 0 \\
 & $b_{\mu}$ & Dilatation (D) & dilatational gauge field & 1 & 0 & 0 \\
 & $V_{\mu}{}^i{}_j$ & SU(4) (V) & $\mathrm{SU}(4)$ gauge field  & 15 & 0 & 0 \\
 & & &  $V_{\mu}{}_i{}^j\equiv(V_{\mu}{}^i{}_j)^\ast=-V_\mu{}^j{}_i$  & & & \\
 & & &  $V_\mu{}^i{}_i=0$ & & &\\
 & $f_{\mu}{}^a$ & Conformal boosts (K) & K-gauge field & 1 & 1 & 0 \\
 & $a_\mu$ & $\mathrm{U}(1)$ & $\mathrm{U}(1)$ gauge field & 1 & 0 & 0 \\
 \hline
\multirow{3}{*}{Fermions} 
 & $\phi_\mu{}^i$ & S-supersymmetry (S)& S-gauge field & $4$ & $\tfrac12$ & $-\frac{1}{2}$ \\
 & & & $\ga_5\,\phi_{\mu}{}^i=-\phi_{\mu}{}^i$ & & &\\
& $\psi_{\mu}{}^i$ & Q-supersymmetry (Q) & gravitino; $\ga_5\,\psi_{\mu}^i=\psi_\mu^i$ & 4 & $-\frac12$ & $-\frac12$ \\
 \hline
 \hline
\end{tabular}
\end{center}
\end{table}

The matter fields of the theory consist of three types of scalar fields $\phi_\alpha, E_{ij}, D^{ij}{}_{kl}$, an antisymmetric tensor $T_{ab}{}^{ij}$ and two spin-1/2 fermions $\Lambda_i, \chi^{ij}{}_k$. We list them in table \ref{table:matter-fields} with their various algebraic properties, and their representation assignments. The rigid $\mathrm{SU}(1,1)$ indices are denoted by $\alpha,\beta=1,2$.

 \begin{table}[h!]
  \caption{Matter fields of $N=4$ conformal supergravity}\label{table:matter-fields}
  \begin{center}
 \begin{tabular}{ c | c l c c c }
 \hline 
 \hline
   & Field & Restrictions & $\mathrm{SU}(4)$ & $w$ & $c$ \\
 \hline
 \multirow{7}{*}{Bosons} & $\phi_{\alpha}$  & $\phi^1=\phi_1^\ast$, $\phi^2=-\phi_2^\ast$ & 1 & 0 & $-1$ \\
 & $E_{ij}$  & $E_{ij}=E_{ji}$ & 10 & 1 & $-1$ \\
 &$T_{ab}{}^{ij}$  & $\tfrac12\varepsilon_{ab}{}^{cd}T_{cd}{}^{ij}=-T_{ab}{}^{ij}$ & 6 & 1 & $-1$\\
 & &  $T_{ab}{}^{ij}=-T_{ab}{}^{ji}$ & & & \\
 & $D^{ij}{}_{kl}$  & $D^{ij}{}_{kl}=\tfrac14 \varepsilon^{ijmn}\varepsilon_{klpq}D^{pq}{}_{mn}$ & $20^\prime$ & 2 & $0$ \\
  & &  $D_{kl}{}^{ij}\equiv(D^{kl}{}_{ij})^\ast=D^{ij}{}_{kl}$ & & & \\
 &  &  $D^{ij}{}_{kj}=0$ & & & \\
 \hline
\multirow{3}{*}{Fermions} 
& $\Lambda_i$  & $\ga_5\Lambda_i=\Lambda_i$ & 4 & $\tfrac{1}{2}$ & $-\tfrac32$ \\
& $\chi^{ij}{}_k$  & $\ga_5\chi^{ij}{}_k=\chi^{ij}{}_k$; $\chi^{ij}{}_k=-\chi^{ji}{}_k$ & 20 & $\tfrac32$ & $-\tfrac12$ \\
 & &  $\chi^{ij}{}_j=0$ & & & \\
 \hline
 \hline
\end{tabular}
\end{center}
\end{table}

An element of $\mathrm{SU}(1,1)$ can be written in terms of the doublet of complex scalars $\phi_\alpha$ which satisfies
\begin{equation}
  \label{eq:SU(11)-inv-cond}
\phi^\alpha\phi_\alpha=1\,,
 \end{equation}
where $\phi^\alpha\equiv\eta^{\alpha\beta}\phi_\beta^\ast$ with $\eta^{\alpha\beta}=\text{diag}(+1,-1)$.  Therefore, due to the presence of the local $\mathrm{U}(1)$, the scalars parametrise an $\mathrm{SU}(1,1)/\mathrm{U}(1)$ coset.

Just as in ordinary gravity where the spin connection is a composite field, the gauge fields $\omega_\mu{}^{ab}$, $f_\mu{}^a$ and $\phi_\mu{}^i$ are expressed in terms of the other ones through a set of conventional constraints on the superconformal curvatures
\begin{align}
\label{eq:cons-curv1}
R(P)_{\mu\nu}{}^a=\,&0\nonumber\,,\\
R(M)_{\mu\nu}{\!}^{ab}e^\nu{}_b=\,&0\nonumber\,,\\
\ga^\mu R(Q)_{\mu\nu}{}^i=\,&0\,.
\end{align}
The $\mathrm{U}(1)$ gauge field $a_\mu$ is also composite and solves the supercovariant constraint
\begin{equation}
\phi^\alpha D_\mu\phi_\alpha=\,-\tfrac{1}{4}\bar\Lambda^i\ga_\mu\Lambda_i\,.\label{eq:U(1)-gaugefield}
 \end{equation}
The derivative $D_\mu$ is covariant with respect to all the gauge symmetries. By making use of the Bianchi identities for the curvatures, the constraints \eqref{eq:cons-curv1} lead to an additional set of identities which are summarised in appendix \ref{App:trans-cur}. The explicit expressions of $R(Q)_{\mu\nu}{\!}^i$ and $R(S)_{\mu\nu}{\!}^i$ are given in appendix \ref{App:trans-cur}. We refer to \cite{Bergshoeff:1980is} for the other ones. 

The independent fields of tables \ref{table:gaugefields} and \ref{table:matter-fields} constitute the full Weyl supermultiplet of $N=4$ conformal supergravity which contains $128+128$ off-shell degrees of freedom. The non-linear superconformal transformation rules of the fields were derived in \cite{Bergshoeff:1980is}. The Q-supersymmetry transformations\footnote{We employed Pauli-K\"all\'en conventions where $x^\alpha$ equals $\mathrm{i}x^0$ for $\alpha=1$, so that all gamma matrices are hermitian.} of the gauge fields read
 \begin{align}
 \label{eq:Q-trans-gauge}
  \delta_Q e_\mu{}^a=\,&\bar\epsilon^i\gamma^a\psi_{\mu i} 
   +\text{h.c.}\,,\nonumber\\ 
   \delta_Q \psi_\mu{}^i=\,&2\,\mathcal{D}_\mu\epsilon^i
   -\tfrac12\gamma^{ab}T_{ab}{\!}^{ij}\gamma_\mu\epsilon_j+\varepsilon^{ijkl}\,
   \bar\psi_{\mu j} \epsilon_k\, \Lambda_l \,,\nonumber\\ 
   \delta_Q b_\mu=\,&\tfrac12\bar\epsilon^i\phi_{\mu i}+\text{h.c.}\,, \nonumber\\ 
   \delta_Q V_{\mu}{}^i{}_j=\,& \bar\epsilon^i\phi_{\mu
     j}+\bar\epsilon^k\gamma_\mu\chi^i{}_{kj}
   -\tfrac12\varepsilon_{jkmn}E^{ik}\,\bar\epsilon^m\psi_{\mu}{}^n
   -\tfrac16E^{ik}\bar\epsilon_j\gamma_\mu\Lambda_k \nonumber\\
   &\, +\tfrac14\varepsilon^{iklm}\,T^{ab}{\!}_{lj}\,
   \bar\epsilon_{k}\gamma_{ab} \gamma_\mu\Lambda_m
   +\tfrac13\bar\epsilon^i\gamma_\mu
   {\Slash{P}}\Lambda_j  \nonumber\\
   &\, -\tfrac14\varepsilon^{iklp} \varepsilon_{jmnp}\,
   \bar\epsilon^m\gamma_a\psi_{\mu k} \,
   \bar\Lambda_l\gamma^a\Lambda^n -(\text{h.c.; traceless})\,,\nonumber\\
      \delta_Q  a_\mu=\,& 
   \tfrac12\mathrm{i} \bar\epsilon_i\gamma_\mu\bar{\Slash{P}}
   \Lambda^i
   +\tfrac14\mathrm{i} E_{ij}\,\bar\Lambda^i\gamma_\mu\epsilon^j
   +\tfrac18 \mathrm{i}  \varepsilon_{ijkl}\, T_{ab}{\!}^{kl}
   \,\bar\Lambda^i\gamma_\mu\gamma^{ab}\epsilon^j
   \nonumber\\
   &\,-\tfrac14\mathrm{i} (\bar\Lambda^i\gamma_a\Lambda_j
   -\delta^i{\!}_j\,\bar\Lambda^k\gamma_a\Lambda_k)
   \,\bar\epsilon_i\gamma^a\psi_\mu{}^j
   +  \text{h.c.}\,,\nonumber\\
   \delta_Q \omega_\mu{\!}^{ab}=\,& 
   -\tfrac12 \bar\epsilon^i\gamma^{ab}\phi_{\mu i}
   +\bar\epsilon^i\gamma_\mu R (Q)^{ab}{}_{i}
   -2\, T^{ab}{\!}_{ij} \,\bar\epsilon^i\psi_\mu{}^j+\text{h.c.}\,,\nonumber\\ 
   \delta_Q f_\mu{}^a=\,& -\tfrac18 e_{\mu b} \,\varepsilon^{abcd}\,
   \bar\epsilon_i R(S)_{cd}{}^i
   -\bar\epsilon_i\gamma_\mu D_b R(Q)^{ab\,i} 
   -2\,T_{\mu b}{}^{ij}\,\bar\epsilon_i R (Q)^{ab}{\!}_j\nonumber\\
   &\,+\text{h.c.}+[\text{terms }\propto \psi_\mu]\,,\nonumber\\
   \delta_Q \phi_{\mu}{}^i=\,&-2f_\mu{}^a \gamma_a\epsilon^i +\tfrac14
   T_{ab}{\!}^{ij}\,T^{cd}{\!}_{jk}\,\gamma_{cd}
   \gamma_{\mu}\gamma^{ab}\epsilon^k \nonumber\\
   &\,+\tfrac16\big[\gamma_\mu\gamma^{ab} -3\,
   \gamma^{ab}\gamma_\mu\big]\big[R(V)_{ab}{}^{i}{}_j\, \epsilon^j +\tfrac12
   \mathrm{i} F_{ab}\epsilon^i+
   \tfrac12 D_a T_{cd}{\!}^{ij} \gamma^{cd}\gamma_b\epsilon_j\big]
   \nonumber\\ 
   &\,+[\text{terms }\propto\psi_\mu] \,,
   \end{align}
   while for the matter fields we have
   \begin{align}
   \label{eq:Q-trans-matter}
   \delta_Q \phi_\alpha=&\,-\eb^i\Lambda_i\varepsilon_{\alpha\beta}\phi^\beta\,,\nonumber\\
     \delta_Q \bar P_a=&\, -\bar\epsilon^iD_a\Lambda_i
   -\tfrac14\bar\Lambda_i\gamma^{bc} T_{bc}{\!}^{ij}\gamma_a\epsilon_i 
   -\tfrac12\bar\epsilon^i\Lambda_i\,\bar\Lambda^j\gamma_a\Lambda_j\,,
   \nonumber\\[1mm]
   \delta_Q \Lambda_i=\,&-2\,\bar{\Slash{P}}\epsilon_i
   +E_{ij}\epsilon^j+\tfrac12\varepsilon_{ijkl}\,T_{bc}{\!}^{kl} 
   \gamma^{bc} \,\epsilon^j \,, \nonumber\\[1mm]
   \delta_Q E_{ij}=&\, 2\, \bar\epsilon_{(i} \Slash{D}\Lambda_{j)}
   -2\,\bar\epsilon^k\chi^{mn}{\!}_{(i} \,\varepsilon_{j)kmn}
   -\bar\Lambda_i\Lambda_j\,\bar\epsilon_k\Lambda^k
   +2\,\bar\Lambda_k\Lambda_{(i}\,\bar\epsilon_{j)}\Lambda^k \,,
   \nonumber\\[1mm]   
   \delta_Q
   T_{ab}{\!}^{ij}=\,& 2\,\bar\epsilon^{[i}R (Q)_{ab}{}^{j]}
   +\tfrac12\bar\epsilon^k\gamma_{ab}\chi^{ij}{}_{k}
   +\tfrac14\varepsilon^{ijkl}\,
   \bar\epsilon_{k}\gamma^c\gamma_{ab}D_c\Lambda_l
   -\tfrac1{6} E^{k[i}\,\bar\epsilon^{j]}\gamma_{ab}\Lambda_k
   +\tfrac13\bar\epsilon^{[i}\gamma_{ab} {\Slash{\bar P}}\Lambda^{j]}\,,
   \nonumber\\[1mm]  
   \delta_Q \chi^{ij}{\!}_k=\,&-\tfrac12\gamma^{ab}
   \Slash{D}T_{ab}{\!}^{ij}\epsilon_k
   -\gamma^{ab} R(V)_{ab}{}^{[i}{}_k \,\epsilon^{j]}
   -\tfrac12\varepsilon^{ijlm}\,\Slash{D} E_{kl}\,\epsilon_m
   +D^{ij}{\!}_{kl}\, \epsilon^l\nonumber\\
   &-\tfrac1{6}\varepsilon_{klmn}E^{l[i}\ga^{ab}\big[T_{ab}{\!}^{j]n}\epsilon^m+T_{ab}{\!}^{mn}\epsilon^{j]}\big]
   +\tfrac12 E_{kl}\,E^{l[i}\,\epsilon^{j]}
   -\tfrac12 \varepsilon^{ijlm}\bar{\Slash{P}}\gamma_{ab}
   T^{ab}{\!}_{kl}\,\epsilon_m\nonumber\\   
   &+\tfrac14 \gamma^a\epsilon_n \big[2\,\varepsilon^{ijln}\bar\chi^{m}{}_{lk}
   -\varepsilon^{ijlm}\bar\chi^{n}{}_{lk}\big]\gamma_a\Lambda_m
   +\tfrac14 \epsilon^{[i} \big[2\,\bar\Lambda^{j]}\Slash{D}\Lambda_k
   +\bar\Lambda_k\Slash{D}\Lambda^{j]}\big]\nonumber\\   
   &-\tfrac14\gamma^{ab}\epsilon^{[i}
   \big[2\, \bar\Lambda^{j]}\gamma_aD_{b}\Lambda_k
   - \bar\Lambda_k\gamma_a D_b\Lambda^{j]}\big]
   -\tfrac{5}{12}\varepsilon^{ijlm}\Lambda_m\,\bar\epsilon_l
   \big[E_{kn}\Lambda^n-2\,{\Slash{P}} \Lambda_k\big]\nonumber\\ 
   &+\tfrac{1}{12}\varepsilon^{ijlm} \Lambda_m\,\bar\epsilon_k\,
   \big[E_{ln}\Lambda^n  -2\,{\Slash{P}}\Lambda_l\big]
   -\tfrac12\gamma^{ab}T_{ab}{\!}^{ij} \gamma^c \epsilon_{[k}\,
   \bar\Lambda^l\gamma_c\Lambda_{l]} \nonumber\\
   &\,-\tfrac12\gamma^{ab} T_{ab}{\!}^{l[i} \gamma^c\epsilon_{[k}\,
   \bar\Lambda^{j]}\gamma_c\Lambda_{l]} 
    +\tfrac12\epsilon^{[i}\bar\Lambda^{j]}\Lambda^m\,
   \bar\Lambda_{k}\Lambda_m 
   -(\text{traces})\,,\nonumber\\[1mm]
   \delta_Q D^{ij}{\!}_{kl}=\,& -4
   \bar\epsilon^{[i}\Slash{D}\chi^{j]}{}_{kl}
   +\varepsilon_{klmn}\,\bar\epsilon^{[i}\big[-2\,E^{j]p}\chi^{mn}{\!}_p
   +\tfrac12\gamma^{ab} T_{ab}{\!}^{mn}
   \stackrel{\leftrightarrow}{\Slash{D}}\Lambda^{j]}
   +\tfrac13 E^{j]m}E^{np}\Lambda_p\nonumber\\
   &\qquad -\tfrac23 {\Slash{\bar P}} \Lambda^mE^{j]n} +\tfrac12
   \gamma^{ab} T_{ab}{\!}^{mn}\Lambda_p\,
   \bar\Lambda^{j]}\Lambda^p\big]  \nonumber\\
   &\, + \bar\epsilon^{[i}\big[2\, \gamma^a\chi^m{}_{kl}\,
   \bar\Lambda^{j]}\gamma_a\Lambda_m + 2\,
   {\Slash{\bar P}}\,\gamma_{ab}T^{ab}{\!}_{kl}\,\Lambda^{j]}
   +\tfrac23\Lambda_{[k}E_{l]m}\,\bar\Lambda^{j]}\Lambda^m
   +\tfrac1{6}\gamma^{ab}{\Slash{P}}\Lambda^{j]}\,
   \bar\Lambda_k\gamma_{ab}\Lambda_l\big] \nonumber\\
   &\, \qquad + \varepsilon^{ijmn}\,
   \bar\epsilon^p\,T^{ab}{\!}_{kl}\big[2\,T_{ab\,np}\Lambda_m+ T_{ab\,mn}\Lambda_{p}\big]
    +(\text{h.c.; traceless}) \,.
  \end{align}
where $\epsilon^i$ is the Q-supersymmetry parameter and where $\mathcal{D}_\mu$ is covariant with respect to the all the bosonic symmetries except the conformal boosts. For instance, we have
  \begin{align}
  \mathcal{D}_\mu \epsilon^i =&\, \big[\partial_\mu -\tfrac14
  \omega_\mu{\!}^{ab} \gamma_{ab} +\tfrac12( b_\mu
  +\mathrm{i} a_\mu)  \big]\epsilon^i  - V_\mu{}^i{}_j\,\epsilon^j \,,
  \nonumber\\ 
  \mathcal{D}_\mu \eta^i =&\, \big[\partial_\mu -\tfrac14
  \omega_\mu{\!}^{ab} \gamma_{ab} - \tfrac12( b_\mu - \mathrm{i} a_\mu)
  \big]\eta^i - V_\mu{}^i{}_j\,\eta^j \,,
\end{align}
where we introduced the S-supersymmetry parameter $\eta^i$. Note that, contrary to \cite{Bergshoeff:1980is}, the $\mathrm{U}(1)$ gauge field is real. 

In \eqref{eq:Q-trans-gauge}, we introduced the supercovariant $\mathrm{U}(1)$ field strength $F_{\mu\nu}$ and the complex vector $P_\mu$ 
\begin{align}
\label{eq:P-def}
P_\mu=\,&\varepsilon_{\alpha\beta}\phi^\alpha D_\mu\phi^\beta\,,\nonumber\\
\bar P_\mu=\,&-\varepsilon^{\alpha\beta}\phi_\alpha D_\mu\phi_\beta\,,
\end{align}
with $\varepsilon_{12}=\varepsilon^{12}=+1$. The S-supersymmetry transformations of the fields are
\begin{align}
    \label{eq:S-trans}
      \delta_S e_\mu{}^a=\,&0\,,\nonumber\\
    \delta_S \psi_{\mu}{}^i=\,&-\gamma_\mu\eta^i\,\nonumber\\
    \delta_S b_\mu=\,&
    -\tfrac12\bar\psi_\mu{\!}^i\,\eta_i+\text{h.c.}\,,\nonumber\\ 
    \delta_S V_\mu{}^i{\!}_j=\,&-\big[\bar\psi_\mu{\!}^i\eta_j
    -\tfrac14\delta^i{\!}_j\,\bar\psi_\mu{\!}^k\eta_k\big]-\text{h.c.}\,,\nonumber\\ 
    \delta_S a_\mu=\,&0\,,\nonumber\\
    \delta_S \omega_{\mu}{}^{ab}=\,&\tfrac12\bar\psi_\mu{}^i\ga^{ab}\eta_i+\text{h.c.}\,,\nonumber\\
\delta_S f_{\mu}{}^a=\,&\tfrac12\bar\eta_i\ga^a\phi_\mu{}^i-\tfrac14\bar\eta_i R(Q)_{\mu}{}^{a\,i}+\tfrac{1}{12}\bar\eta_i\ga^{bc} T_{bc}{\!}^{ij}\ga^a\psi_{\mu j}+\text{h.c.}\,,\nonumber\\
\delta_S \phi_{\mu}{}^i=\,&2\,\mathcal{D}_\mu\eta^i-\tfrac16	\ga_\mu\ga^{ab} T_{ab}{\!}^{ij}\eta_j+\tfrac12\varepsilon^{ijkl}\bar\eta_k\Lambda_l\psi_{\mu j}\,,\nonumber\\
\delta_S \phi_\alpha=\,& 0 \nonumber\\
    \delta_S \bar{P}_a=\,&-\tfrac12\bar\eta^i\gamma_a\Lambda_i\,,\nonumber\\
    \delta_S \Lambda_i=\,&0\,,\nonumber\\
    \delta_S E_{ij}=\,& 2\,\bar\eta_{(i}\Lambda_{j)}\,,\nonumber\\
    \delta_S T_{ab}{}^{ij}=\,&-\tfrac14\varepsilon^{ijkl}\,
    \bar\eta_k\gamma_{ab}\Lambda_l\,,
    \nonumber\\
    \delta_S \chi^{ij}{}_k=\,&\tfrac12 
    T_{ab}{\!}^{ij}\,\gamma^{ab}\eta_k\, 
    +\tfrac{2}{3}\delta^{[i}_{k}\,T_{ab}{\!}^{j]l}\, \gamma^{ab}
    \eta_l-\tfrac12\varepsilon^{ijlm}E_{kl}\,\eta_m  
    -\tfrac14\bar\Lambda_k\gamma^a\Lambda^{[i}\gamma_a\eta^{j]}\nonumber\\
    &+\tfrac{1}{12}\delta^{[i}_{k}\big[\bar\Lambda_l\gamma^a\Lambda^l\gamma_a\eta^{j]}
    -\bar\Lambda_l\gamma^a\Lambda^{j]}\,\gamma_a\eta^{l}\big]\,,\nonumber\\
    \delta_S D^{ij}{}_{kl}=\,&0\,.
  \end{align}

As is clear from \eqref{eq:Q-trans-gauge}, \eqref{eq:Q-trans-matter} and \eqref{eq:S-trans}, the coset space sector of the theory can be entirely described in terms of $P_\mu$ and $F_{\mu\nu}$. In what follows, we will make use of these $\mathrm{SU}(1,1)$ invariant quantities rather than the scalars $\phi_\alpha$. Note also that $P_a$ has Weyl weight $w=1$ and is invariant under K-transformations. We finally present several identities which will be useful in the next sections. Using \eqref{eq:U(1)-gaugefield} and \eqref{eq:P-def}, one can respectively derive 
\begin{align}
\label{eq:id-scalar1}
\varepsilon_{\beta\gamma}D_a\phi^\beta D_b\phi^\gamma=\,&2\,\phi_\alpha D_{[a}\phi^\alpha\varepsilon_{\beta\gamma}\phi^\beta D_{b]}\phi^\gamma=\tfrac12\bar\Lambda^i\ga_{[a}\Lambda_i  P_{b]}\,,\nonumber\\
D_a\phi^\alpha D_b\phi_\alpha=\,&-P_a\bar P_b-\tfrac{1}{16}\bar\Lambda^i\ga_a\Lambda_i\bar\Lambda^j\ga_b\Lambda_j\,.
\end{align}
It follows that
\begin{align}
F_{ab}=&\,2\mathrm{i}\bar{P}_{[a} P_{b]}-\tfrac{1}{2}\mathrm{i}\big[\bar\Lambda^i\ga_{[a}D_{b]}\Lambda_i-\text{h.c.}\big]\label{eq:id-scalar2}\,,\\
D_{[a} \bar{P}_{b]}=&\,\tfrac12\bar\Lambda_i\ga_{[a}\Lambda^i  \bar{P}_{b]}+\tfrac14\bar\Lambda_iR(Q)_{ab}{}^i\,,\label{eq:id-scalar3}
\end{align}
which are the supersymmetric generalisations of the Maurer-Cartan equations associated with the coset space $\mathrm{SU}(1,1)/\mathrm{U}(1)$. 

\setcounter{equation}{0}

\section{The quadratic action} \label{sec:quadratic-lagrangian}
In this section, we present the part of the action which is quadratic in the fields. It will be the starting point for the iterative procedure presented in section \ref{sec:strategy}, which we will use to generate terms of higher-order in the fields. The action will be constructed such that all the derivatives and curvatures that appear are fully supercovariantized with respect to all the gauge transformations (bosonic as well as fermionic). Hence, we must insist that, throughout the paper, our counting of the fields always excludes the gauge fields which are implicitly contained within the supercovariant derivatives and curvatures.

The quadratic Lagrangian of $N=4$ conformal supergravity reads
\begin{align}
\label{eq:quad-action}
e^{-1}\LL_{\text{Q}} = \,&\,\tfrac{1}{2}{R}(M)^{abcd}R(M)^{-}_{abcd}+{R}(V)^{ab\,i}{}_{j}{R}(V)^{-~j}_{ab~~i}\nonumber\\
& - 4\,T_{ab}^{~~ij}D^{a}D_{c}T^{cb}{\!}_{ij}+\tfrac{1}{4}E_{ij}D^2E^{ij}+{1\over 8}D_{ij}^{~~kl}D_{kl}^{~~ij}\nonumber\\
&-2\,\bar{P}^{a}\big[D_{a}D^{b}P_{b}+D^{2}P_{a}\big]-2\,D^{a}P^{b}D_{a}\bar P_{b}-D^{a}P_{a}D^{b}\bar P_{b}\nonumber\\
&+\bar{R}(Q)_{ab}{}^{i}R(S)^{ab}{}_{i}-\bar{\chi}^{ij}{}_{k}\Dslash\chi^{k}_{~ij}-\tfrac{1}{2}\bar{\Lambda}_{i}\left(D^{2}\Dslash+\Dslash D^{2}-\Dslash^{3}\right)\Lambda^{i}\,+\thc\,,
\end{align}
with $e=\text{det}[e_\mu{}^a]$ and where the (anti)self-dual part of a generic second rank tensor $R_{ab}$ is defined as $R^{\pm}_{ab}=\tfrac12[R_{ab}\pm\tfrac12\varepsilon_{abcd}R^{cd}]$.
The expression \eqref{eq:quad-action} corresponds to the real part of the chiral invariant of the linearized theory given in \cite{Bergshoeff:1980is}. The imaginary part of the chiral invariant is a total derivative.

The structures of the quadratic terms are uniquely fixed by requiring invariance under $\mathrm{U}(1)$, $\mathrm{SU}(4)$ and Lorentz symmetry, while the number of derivatives in each term is fixed by Weyl invariance. At the level of the action, the derivatives can be moved around using integration by parts at the expense of higher-order terms in the fermions. However, requiring K-invariance (i.e. under conformal boosts) fixes the position of the derivatives. Under these conditions, the quadratic terms for the fields $E^{ij}, T_{ab}{\!}^{ij}$and $\Lambda_i$ are uniquely determined. The case of the vectors $P_\mu$ is more subtle and will be discussed below.

The relative coefficients between the different quadratic terms are fixed by requiring Q-supersymmetry invariance at quadratic order in the fields. The K-invariance of the quadratic terms involving the vectors $P_\mu$ is not straightforward. Out of the four possible terms, all appearing in the Lagrangian \eqref{eq:quad-action}, none is K-invariant. The two terms in which both derivatives act on the same field should not be treated as independent. Indeed, only their sum is relevant at quadratic order since their difference
\begin{equation}
\label{id-dif}
D^2 P_a-D_aD^b P_b=\,D^bD_{[b} P_{a]}+[D^b,D_a] P_b\,,
\end{equation}
is of higher-order in the fields due to \eqref{eq:id-scalar3}. An arbitrary combination of the remaining three independent quadratic terms is generically not K-invariant. However, when considering the unique combination appearing in \refb{eq:quad-action}, one finds that it is K-invariant up to a term of higher-order in the fields
\begin{align}
\label{eq:K-var-scal-quad}
\delta_K \big[2\,\bar{P}^{a}&\big(D_{a}D^{b} {P}_{b}+D^{2} P_{a}\big)\nonumber\\
&+2\,D_{a}P_{b}D^{a}\bar P^{b}+D_{a}P^{a}D_{b}\bar P^{b}+\thc\big]=\,4\,\Lambda^K_{a}\bar{P}_{b}D^{[b} P^{a]}+\thc \,.
\end{align}
Here $\Lambda_a^K$ is the K-transformation parameter. We should emphasise that, at this point, requiring K-invariance of each of the supercovariant terms in the Lagrangian is not necessary. The advantage of imposing such a condition already at the level of the quadratic action is that terms with an explicit K-gauge field $f_{\mu}^{~a}$ will not have to be introduced when deriving the interaction terms. This will be explained in section \ref{sec:strategy}.

Finally, it is important to emphasize that in this paper, we will exclusively consider the real part of the chiral invariant. Without this reality condition, the K-variation of the kinetic terms for $P_\mu$ is not of higher-order in the fields anymore  and consequently, one is forced to introduce explicit K-gauge fields.

\setcounter{equation}{0}

\section{Building up higher-order terms}
\label{sec:strategy}

In this section, we present the iterative procedure used to construct the supersymmetric completion of the quadratic Lagrangian \eqref{eq:quad-action}. The  non-linearity of the supersymmetry transformations rules will require us to add successive layers of terms of higher-order in the fields to the Lagrangian. The higher-order terms will be chosen such that their supersymmetry variations precisely cancel against the variations of the pre-existing lower-order terms. Ultimately, this program terminates when all the necessary terms have been added such that the Lagrangian is fully invariant under supersymmetry. Requiring Q-supersymmetry invariance turns out to be enough to ensure invariance under all the symmetries of $N=4$ conformal supergravity. This is due to the specific superalgebra obeyed by the different generators \cite{Bergshoeff:1980is}. Indeed, the commutator of two infinitesimal Q-supersymmetry transformations yields the full set of superconformal transformations including the $\mathrm{U}(1)$ transformation.

\subsection{Structure of the full Lagrangian}\label{subset:structure}

This supersymmetrization procedure is unambiguous, yet lengthy, and provided sufficient computational efforts are invested it is guaranteed to give the full off-shell superconformal invariant. In practice however, the computation rapidly becomes unmanageable due to the rich field content and the non-linearity of the transformation rules. Therefore it becomes essential to systematise the work by making use of certain structure patterns appearing 
in the computation. Hence, we argue\footnote{This is inspired by the approaches of \cite{Castellani1991:bg,Gates1991:bg}.} that the full Lagrangian can be written in the following form
\begin{equation}
\label{eq:Lagrangian-structure}
\LL=\LL_{0}+\psi \LL_{\psi} +\phi\LL_{\phi}+\psi^2 \LL_{\psi^2}+\psi\phi\LL_{\psi\phi}+ \phi^2\LL_{\phi^2}+\psi^3 \LL_{\psi^3} +\psi^2 \phi \LL_{\psi^2\phi}+\psi^{4}\LL_{\psi^4}\,,
\end{equation}
where here, $\psi$ and $\phi$ schematically denote the gravitino and the S-gauge field, respectively. The quantities $\LL_0,\LL_\psi,\LL_\phi,\LL_{\psi^2},\LL_{\phi^2},\LL_{\psi\phi}$, $\LL_{\psi^3}$, $\LL_{\psi^2 \phi}$, $\LL_{\psi^4}$ only depend on supercovariant fields, i.e. matter fields, supercovariant curvatures and their supercovariant derivatives. Note that the terms of lowest-order in the fields in $\LL_0$ correspond to the quadratic Lagrangian \eqref{eq:quad-action}. Consequently, the other supercovariant quantities in \eqref{eq:Lagrangian-structure} are at least of quadratic order in the fields.

The expression \eqref{eq:Lagrangian-structure} only contains terms up to four explicit gauge fields ($\psi$ or $\phi$). This can be understood as follows. Under an infinitesimal Q-supersymmetry variation (Q-variation), a gravitino transforms into the gradient of the Q-supersymmetry parameter. In order for this variation to be subsequently canceled, it first has to be integrated by parts such that when the derivative hits any of the other explicit gauge fields, it yields a curvature (Q or S). This requires the explicit gauge fields to appear fully anti-symmetrized in their vector indices and therefore rules out the possibility of terms with more than four explicit gauge fields. The same reasoning holds for an infinitesimal S-supersymmetry variation acting on $\phi$. However, for our current analysis the terms with more than two explicit gauge fields are not required since we are only looking to construct the Lagrangian up to quadratic order in the fermion fields. We will therefore not attempt to derive them explicitly.

 The Weyl weights of $\psi$ and $\phi$ restrict the order of the possible terms appearing in the various quantities $\LL_{0},\LL_{\psi},\ldots$. For instance, based on the fact that the field $\Lambda_i$ has the lowest Weyl weight, one expects the terms of $\LL_0$ to be at most of eigth-order in $\Lambda_i$ without any derivatives. Weyl invariance also rules out  terms with more than two S-gauge fields. Furthermore, terms with two $\phi$'s and one $\psi$ do not appear in \eqref{eq:Lagrangian-structure} as the Weyl weight of their associated supercovariant factor does not allow for more than one covariant field. For the same reason, terms with three $\psi$'s and one $\phi$ are not present. Because of the Weyl weight of $\phi$, the term $\phi^2\LL_{\phi^2}$ will be of higher-order in the fermion fields\footnote{If $\phi^2\LL_{\phi^2}$ would contain terms which are quadratic in the fermions, then for our purposes $\LL_{\phi^2}$ would have to be purely bosonic. This possibility is again ruled out by the Weyl weight of the bosonic fields.}.

Finally, in order to write the full Lagrangian as in \eqref{eq:Lagrangian-structure}, we assumed that there are no terms containing explicit K-gauge fields. Because of its Weyl weight, the K-gauge field $f$ could only schematically appear within terms of the form $f\LL_f$ and $\psi f\LL_{\psi f}$ where $\LL_f,\LL_{\psi f}$ are supercovariant. However, some parts of the S-supersymmetry variations of these two terms would necessarily have to cancel against each other, and consequently the absence of one implies the absence of the other. Since the first one could only arise to compensate for the lack of K-invariance of $\LL_0$, it means that a K-invariant $\LL_0$ prohibits the appearance of explicit K-gauge fields throughout the full Lagrangian. In section \ref{sec:quadratic-lagrangian}, we have written the quadratic part of $\LL_0$ in such a way that it is K-invariant at quadratic order in the fields. As will be clear from our results in section \ref{sec:results}, the completion of $\LL_0$ to higher-order in the fields is K-invariant and therefore there will be no deviation from the structure \eqref{eq:Lagrangian-structure}.

Finally, it is clear that the expression \eqref{eq:Lagrangian-structure} cannot capture accurately the structure of the full chiral invariant. Indeed, as was discussed in section \ref{sec:quadratic-lagrangian}, the latter involves explicit K-gauge fields.

\subsection{Constructing the interaction terms}

In this subsection, we outline the iterative procedure used to construct the various supercovariant quantities appearing in the schematic expression \eqref{eq:Lagrangian-structure} of the full $N=4$ conformal supergravity Lagrangian. To this purpose, let us first write a part of \eqref{eq:Lagrangian-structure} with explicit indices
\begin{align}\label{eq:Lagrangian-exact}
\LL=&\,\LL_{0}+\left[{1\over 2}\bar\psi_{a}{\!}^{i} \LL_{\scriptscriptstyle{\psi}}{\!}^a{\!}_i +{1\over 2}\bar\phi_{a}{}^{i}\LL_{\scriptscriptstyle {\phi}}{\!}^a{\!}_i+ {1\over 4}\bar{\psi}_{b}{}^{i}\LL_{\scriptscriptstyle {\psi^2}}{\!}^{ab}{\!}_{ij}\psi_{a}{}^{j}+{1\over4}\bar{\psi}_{bi}\LL_{\scriptscriptstyle {\psi^2}}{\!}^{ab\,i}{\!}_j\psi_{a}{}^{j}\right.\nonumber \\
&\left. \,\,\,\,\,\,\,\,\,\,\,\,\,\,\;\;\;\;\;\;\;+{1\over 2}\bar{\psi}_{b}{}^{i}\LL_{\scriptscriptstyle {\psi\phi}}{\!}^{ab}{\!}_{ij}\phi_{a}{}^{j}+{1\over2}\bar{\psi}_{bi}\LL_{\scriptscriptstyle {\psi\phi}}{\!}^{ab\,i}{\!}_j\phi_{a}{}^{j} +\thc \right]\,.
\end{align}
Since we are only interested in the Lagrangian up to quadratic order in the fermion fields, we have truncated the full Lagrangian to the above expression. For the same reason, $\LL_0$ is restricted to terms up to quadratic order in the fermions, while $\LL_{\scriptscriptstyle {\psi}}{\!}^a{\!}_i,\LL_{\scriptscriptstyle {\phi}}{\!}^a{\!}_i$ and $\LL_{\scriptscriptstyle {\psi^2}}{\!}^{ab}{\!}_{ij},\LL_{\scriptscriptstyle {\psi^2}}{\!}^{ab\,i}{\!}_j,\LL_{\scriptscriptstyle {\psi\phi}}{\!}^{ab}{\!}_{ij},\LL_{\scriptscriptstyle {\psi\phi}}{\!}^{ab\,i}{\!}_j$ are only linear in the fermions and purely bosonic\footnote{They contain only bosonic fields but they are still matrices in the spinor space.}, respectively.  Note also that, as discussed in section \ref{subset:structure}, the last four quantities are antisymmetric in their vector indices.

In what follows, we will work at specific orders in the supercovariant fields. To this purpose, we define $\LL^{\scriptscriptstyle{(n)}}_0,\LL^{\scriptscriptstyle{(n)}}_{\scriptscriptstyle {\psi}}{}^a{\!}_i, \LL^{\scriptscriptstyle{(n)}}_{\scriptscriptstyle {\phi}}{}^a{\!}_i,\LL^{\scriptscriptstyle{(n)}}_{\scriptscriptstyle {\psi^2}}{}^{ab}{\!}_{ij},\LL^{\scriptscriptstyle{(n)}}_{\scriptscriptstyle {\psi^2}}{}^{ab\,i}{\!}_j$ and $\LL^{\scriptscriptstyle{(n)}}_{\scriptscriptstyle {\psi\phi}}{}^{ab}{\!}_{ij},\LL^{\scriptscriptstyle{(n)}}_{\scriptscriptstyle {\psi\phi}}{}^{ab\,i}{\!}_j$ which contain the terms of order $n$ in the supercovariant fields of the quantities appearing in \eqref{eq:Lagrangian-structure}. They will be constructed by requiring that the various Q-variations of order $n$ vanish. These variations naturally arise from terms of order $n$ in the Lagrangian but also from terms of lowest-order. Therefore, each layer of computation relies on the previous ones. Consequently, all the terms at order $n<m$ have to be constructed before the terms of order $m$. Furthermore, we can systematically restrict ourselves to variations which are linear in the fermion fields since we are only looking to derive the terms in the Lagrangian up to quadratic order in the fermions.

In order to explain how the Q-variations at a specific order cancel against each other, we compute below the Q-variations of the various terms appearing in the Lagrangian at order $n$.  
To this purpose, we introduce the symbols $\delta_{K\vert_{f_a}},\delta_{Q\vert_{\psi_a}}$ and $\delta_{S\vert_{\phi_a}}$ which denote gauge transformations where the parameters are replaced by the associated gauge fields. Additionally, we define $\delta_Q^{\scriptscriptstyle{(cov)}}$ as the supercovariant part of a Q-variation. In what follows, we insist that all the variations which are of cubic order, or more than cubic order, in the fermions (gauge and matter fields) will be suppressed.
\begin{align}
\delta_Q\LL^{\scriptscriptstyle{(n)}}_0&\sim\,[\delta_Qe] e^{-1}\LL^{\scriptscriptstyle{(n)}}_0+e\,\delta_Q[e^{-1}\LL^{\scriptscriptstyle{(n)}}_0]\,,\label{eq:var1}\\[1mm]
\tfrac12\delta_Q[ \bar\phi_{a}{\!}^{i} \LL^{\scriptscriptstyle{(n)}}_{\scriptscriptstyle{\phi}}{}^a{\!}_i+\thc ]&\sim\,f_a{}^b\eb^i\ga_b \LL^{\scriptscriptstyle{(n)}}_{\scriptscriptstyle{\phi}}{}^a{\!}_i+\tfrac12[\delta_Q^{\scriptscriptstyle{(cov)}}\bar\phi_a{\!}^i]\LL^{\scriptscriptstyle{(n)}}_{\scriptscriptstyle{\phi}}{}^a{\!}_i+\tfrac{e}{2} \bar\phi_{a}{\!}^{i} \delta_Q[e^{-1}\LL^{\scriptscriptstyle{(n)}}_{\scriptscriptstyle{\phi}}{}^a{\!}_i ]+\thc\,,\label{eq:var2}\\[1mm]
\tfrac12\delta_Q[ \bar\psi_{a}{\!}^{i} \LL^{\scriptscriptstyle{(n)}}_{\scriptscriptstyle{\psi}}{}^a{}_i +\thc]&\sim\,\mathcal{D}_a\eb^i \LL^{\scriptscriptstyle{(n)}}_{\scriptscriptstyle{\psi}}{}^a{\!}_i- \tfrac14\eb_j\ga_a\ga\cdot T^{ij}\LL^{\scriptscriptstyle{(n)}}_{\scriptscriptstyle{\psi}}{}^a{}_i+\tfrac{e}{2}\bar\psi_{a}{\!}^{i}\delta_Q [e^{-1}\LL^{\scriptscriptstyle{(n)}}_{\scriptscriptstyle{\psi}}{}^a{\!}_i] +\thc\nonumber\\[1mm]
&\sim\,-e\,\eb^iD_a[e^{-1} \LL^{\scriptscriptstyle{(n)}}_{\scriptscriptstyle{\psi}}{}^a{\!}_i] -\eb^i\delta_{K\vert_{f_a}}\LL^{\scriptscriptstyle{(n)}}_{\scriptscriptstyle{\psi}}{}^a{\!}_i -\tfrac{e}{2}\eb^i\delta_{Q\vert_{\psi_a}}[e^{-1}\LL^{\scriptscriptstyle{(n)}}_{\scriptscriptstyle{\psi}}{}^a{\!}_i]\nonumber\\[1mm]
&\;\;\;\;\;-\tfrac{e}{2}\eb^i\delta_{S\vert_{\phi_a}}[e^{-1}\LL^{\scriptscriptstyle{(n)}}_{\scriptscriptstyle{\psi}}{}^a{\!}_i]- \tfrac14\eb_j\ga_a\ga\cdot T^{ij}\LL^{\scriptscriptstyle{(n)}}_{\scriptscriptstyle{\psi}}{}^a{}_i+\tfrac{e}{2}\bar\psi_{a}{\!}^{i}\delta_Q [e^{-1}\LL^{\scriptscriptstyle{(n)}}_{\scriptscriptstyle{\psi}}{}^a{\!}_i]\nonumber\\[1mm]
&\;\;\;\;\;\,+\thc\,,\label{eq:var3}
\end{align}
where in \eqref{eq:var3}, we dropped a total derivative. Note that the term involving the field $T_{ab}{\!}^{ij}$ comes from the covariant part of the variation of $\psi_a{\!}^i$. It will appear similarly in the subsequent variations. We continue with
\begin{align}
\tfrac14\delta_Q[\bar{\psi}_{b}{}^{i}\LL^{\scriptscriptstyle{(n)}}_{\scriptscriptstyle {\psi^2}}{}^{ab}{\!}_{ij}&\psi_{a}{}^{j}+\bar{\psi}_{bi}\LL^{\scriptscriptstyle{(n)}}_{\scriptscriptstyle {\psi^2}}{}^{ab\,i}{\!}_j\psi_{a}{}^{j}+\thc]\nonumber\\[1mm]
&\sim\,[\mathcal{D}_b\eb^i]\LL^{\scriptscriptstyle{(n)}}_{\scriptscriptstyle {\psi^2}}{}^{ab}{\!}_{ij}\psi_{a}{}^{j}+[\mathcal{D}_b\eb^i]\LL^{\scriptscriptstyle{(n)}}_{\scriptscriptstyle {\psi^2}}{}^{ab\,i}{\!}_j\psi_{a}{}^{j}\nonumber\\[1mm]
&\;\;\;\;\;-\tfrac14[\eb_k\ga_b\ga\cdot T^{ik}\LL^{\scriptscriptstyle{(n)}}_{\scriptscriptstyle {\psi^2}}{}^{ab}{\!}_{ij}\psi_{a}{}^{j}+\eb^k\ga_b\ga\cdot T_{ik} \LL^{\scriptscriptstyle{(n)}}_{\scriptscriptstyle {\psi^2}}{}^{ab\,i}{\!}_j\psi_{a}{}^{j}]+\thc\nonumber\\[1mm]
&\sim\,-e\,\eb^iD_b[e^{-1}\LL^{\scriptscriptstyle{(n)}}_{\scriptscriptstyle {\psi^2}}{}^{ab}{\!}_{ij}]\psi_{a}{}^{j}-e\,\eb_iD_b[e^{-1}\LL^{\scriptscriptstyle{(n)}}_{\scriptscriptstyle {\psi^2}}{}^{ab\,i}{\!}_j]\psi_{a}{}^{j}\nonumber\\[1mm]
&\;\;\;\;\;-\eb^i [\delta_{K\vert_{f_b}}\LL^{\scriptscriptstyle{(n)}}_{\scriptscriptstyle {\psi^2}}{}^{ab}{\!}_{ij}]\psi_{a}{}^{j}-\eb_i[\delta_{K\vert_{f_b}}\LL^{\scriptscriptstyle{(n)}}_{\scriptscriptstyle {\psi^2}}{}^{ab\,i}{\!}_j]\psi_{a}{}^{j}\nonumber\\[1mm]
&\;\;\;\;\;-\tfrac12\big[\eb^i\LL^{\scriptscriptstyle{(n)}}_{\scriptscriptstyle {\psi^2}}{}^{ab}{\!}_{ij}+\eb_i\LL^{\scriptscriptstyle{(n)}}_{\scriptscriptstyle {\psi^2}}{}^{ab\,i}{\!}_j\big]\big[\ga_{b}\phi_{a}{}^{j}+R(Q)_{ba}{\!}^j+\tfrac12\ga\cdot T^{jk}\ga_{b}\psi_{ak}\big]\nonumber\\[1mm]
&\;\;\;\;\;-\tfrac14[\eb_k\ga_b\ga\cdot T^{ik}\LL^{\scriptscriptstyle{(n)}}_{\scriptscriptstyle {\psi^2}}{}^{ab}{\!}_{ij}\psi_{a}{}^{j}+\eb^k\ga_b\ga\cdot T_{ik} \LL^{\scriptscriptstyle{(n)}}_{\scriptscriptstyle {\psi^2}}{}^{ab\,i}{\!}_j\psi_{a}{}^{j}]+\thc\,,\label{eq:var4}
\end{align}
where we have again dropped a total derivative. In the sixth line, we have used that $\LL^{\scriptscriptstyle{(n)}}_{\scriptscriptstyle {\psi^2}}{}^{ab}{\!}_{ij},\LL^{\scriptscriptstyle{(n)}}_{\scriptscriptstyle {\psi^2}}{}^{ab\,i}{\!}_j$ are antisymmetric in their vector indices and we have rewritten the curl of the gravitino  making use of the explicit expression of $R(Q)_{ab}{\!}^i$ given in \eqref{eq:Qcurv}. Finally, we have
\begin{align}
\tfrac12&\delta_Q[\bar{\psi}_{b}{}^{i}\LL^{\scriptscriptstyle{(n)}}_{\scriptscriptstyle {\psi\phi}}{}^{ab}{\!}_{ij}\phi_{a}{}^{j}+\bar{\psi}_{bi}\LL^{\scriptscriptstyle{(n)}}_{\scriptscriptstyle {\psi\phi}}{}^{ab\,i}{\!}_j\phi_{a}{}^{j} +\thc]\nonumber\\[1mm]
&\sim\,[\mathcal{D}_b\eb^{i}]\LL^{\scriptscriptstyle{(n)}}_{\scriptscriptstyle {\psi\phi}}{}^{ab}{\!}_{ij}\phi_{a}{}^{j}+[\mathcal{D}_b\eb_i]\LL^{\scriptscriptstyle{(n)}}_{\scriptscriptstyle {\psi\phi}}{}^{ab\,i}{\!}_j\phi_{a}{}^{j}\nonumber\\[1mm]
&\;\;\;\;\;-\tfrac14[\eb_k\ga_b\ga\cdot T^{ik}\LL^{\scriptscriptstyle{(n)}}_{\scriptscriptstyle {\psi\phi}}{}^{ab}{\!}_{ij}\phi_{a}{}^{j}+\eb^k\ga_b\ga\cdot T_{ik} \LL^{\scriptscriptstyle{(n)}}_{\scriptscriptstyle {\psi\phi}}{}^{ab\,i}{\!}_j\phi_{a}{}^{j}]\nonumber\\[1mm]
&\;\;\;\;\;-\bar{\psi}_{b}{}^{i}\LL^{\scriptscriptstyle{(n)}}_{\scriptscriptstyle {\psi\phi}}{}^{ab}{\!}_{ij}\ga_c\,\epsilon^jf_{b}{\!}^c-\bar{\psi}_{bi}\LL^{\scriptscriptstyle{(n)}}_{\scriptscriptstyle {\psi\phi}}{}^{ab}{\!}_{ij}\ga_c\,\epsilon^jf_{b}{\!}^c+\tfrac12[\bar{\psi}_{b}{}^{i}\LL^{\scriptscriptstyle{(n)}}_{\scriptscriptstyle {\psi\phi}}{}^{ab}{\!}_{ij}+\bar{\psi}_{bi}\LL^{\scriptscriptstyle{(n)}}_{\scriptscriptstyle {\psi\phi}}{}^{ab\,i}{\!}_j]\delta_Q^{\scriptscriptstyle{(cov)}} \phi_a{\!}^j+\thc\nonumber\\[1mm]
&\,\sim-e\,\eb^{i}D_b[e^{-1}\LL^{\scriptscriptstyle{(n)}}_{\scriptscriptstyle {\psi\phi}}{}^{ab}{\!}_{ij}]\phi_{a}{}^{j}-e\,\eb_iD_b[e^{-1}\LL^{\scriptscriptstyle{(n)}}_{\scriptscriptstyle {\psi\phi}}{}^{ab\,i}{\!}_j]\phi_{a}{}^{j}\nonumber\\[1mm]
&\;\;\;\;\;-\tfrac14[\eb_k\ga_b\ga\cdot T^{ik}\LL^{\scriptscriptstyle{(n)}}_{\scriptscriptstyle {\psi^2}}{}^{ab}{\!}_{ij}\phi_{a}{}^{j}+\eb^k\ga_b\ga\cdot T_{ik} \LL^{\scriptscriptstyle{(n)}}_{\scriptscriptstyle {\psi^2}}{}^{ab\,i}{\!}_j\phi_{a}{}^{j}]\nonumber\\[1mm]
&\;\;\;\;\;-\bar{\psi}_{b}{}^{i}\LL^{\scriptscriptstyle{(n)}}_{\scriptscriptstyle {\psi\phi}}{}^{ab}{\!}_{ij}\ga_c\,\epsilon^jf_{b}{\!}^c-\bar{\psi}_{bi}\LL^{\scriptscriptstyle{(n)}}_{\scriptscriptstyle {\psi\phi}}{}^{ab}{\!}_{ij}\ga_c\,\epsilon^jf_{b}{\!}^c+\tfrac12[\bar{\psi}_{b}{}^{i}\LL^{\scriptscriptstyle{(n)}}_{\scriptscriptstyle {\psi\phi}}{}^{ab}{\!}_{ij}+\bar{\psi}_{bi}\LL^{\scriptscriptstyle{(n)}}_{\scriptscriptstyle {\psi\phi}}{}^{ab\,i}{\!}_j]\delta_Q^{\scriptscriptstyle{(cov)}} \phi_a{\!}^j\nonumber\\[1mm]
&\;\;\;\;\;-\tfrac12\big[\eb^i\LL^{\scriptscriptstyle{(n)}}_{\scriptscriptstyle {\psi\phi}}{}^{ab}{\!}_{ij}+\eb_i\LL^{\scriptscriptstyle{(n)}}_{\scriptscriptstyle {\psi\phi}}{}^{ab\,i}{\!}_j\big]\big[R(S)_{ba}{\!}^j-2\ga_c\psi_{a}{\!}^jf_{b}{\!}^c+\tfrac16\ga_{b}\ga\cdot T^{jk}\phi_{ak}-\delta^{\scriptscriptstyle{(cov)}}_{Q\vert_{\psi_{a}}}\phi_{b}{\!}^j\big]+\thc\label{eq:var5}
\end{align}
where after dropping a total derivative, we used in the last line that $\LL^{\scriptscriptstyle{(n)}}_{\scriptscriptstyle {\psi\phi}}{}^{ab}{\!}_{ij},\LL^{\scriptscriptstyle{(n)}}_{\scriptscriptstyle {\psi\phi}}{}^{ab\,i}{\!}_j$ are antisymmetric in their vector indices. This allowed us to rewrite the curl of the S-gauge field through the expression of $R(S)_{ab}{\!}^i$ given in \eqref{eq:Scurv}. Note that we have also used $\delta_K\LL^{\scriptscriptstyle{(n)}}_{\scriptscriptstyle {\psi\phi}}{}^{ab}{\!}_{ij}=\delta_K \LL^{\scriptscriptstyle{(n)}}_{\scriptscriptstyle {\psi\phi}}{}^{ab\,i}{\!}_j=0$. This is because $n\geq 2$, and in our case, $\LL^{\scriptscriptstyle{(n)}}_{\scriptscriptstyle {\psi\phi}}{}^{ab}{\!}_{ij}, \LL^{\scriptscriptstyle{(n)}}_{\scriptscriptstyle {\psi\phi}}{}^{ab\,i}{\!}_j$ are bosonic quantities with Weyl weight 2. 

We now present in detail how the different variations appearing in \eqref{eq:var1}--\eqref{eq:var5} cancel each other out up to order $n$ in the supercovariant fields. The purely supercovariant variations must cancel as
\begin{align}\label{eq:cancel-supercov}
& \tfrac{1}{2} \sum_{k=2}^{n}\delta_Q\left(e^{-1}\LL_{0}^{(k)}\right) -\tfrac{1}{4}\eb_{j}\gamma_{a}\gamma.T^{ij}\sum_{k=2}^{n-1} \left(e^{-1}\LL^{\scriptscriptstyle{(k)}}_{\scriptscriptstyle{\psi}}{}^a{\!}_i\right)+\tfrac{1}{2}\big[\delta_Q^{\scriptscriptstyle{(cov)}}\bar{\phi}_{a}{\!}^{i}\big]\sum_{k=1}^{n-1}\left(e^{-1}\LL^{\scriptscriptstyle{(k)}}_{\scriptscriptstyle{\phi}}{}^a{\!}_i\right) \nonumber \\
&\! -\tfrac{1}{2}\eb^{i}\sum_{k=2}^{n-1}\left(e^{-1}\LL^{\scriptscriptstyle{(k)}}_{\scriptscriptstyle{\psi^2}}{}^{ab}{\!}_{ij}\right){R(Q)}_{ba}{\!}^j-\tfrac{1}{2}\eb_{i}\sum_{k=2}^{n-1}\left(e^{-1}\LL^{\scriptscriptstyle{(k)}}_{\scriptscriptstyle{\psi^2}}{}^{ab}{}^{i}{\!}_{j}\right){R(Q)}_{ba}{\!}^j \nonumber \\
& -\tfrac{1}{2}\eb^{i}\sum_{k=2}^{n-1}\left(e^{-1}\LL^{\scriptscriptstyle{(k)}}_{\scriptscriptstyle{\psi\phi}}{}^{ab}{\!}_{ij}\right){R(S)}_{ba}{\!}^j-\tfrac{1}{2}\eb_{i}\sum_{k=2}^{n-1}\left(e^{-1}\LL^{\scriptscriptstyle{(k)}}_{\scriptscriptstyle{\psi\phi}}{}^{ab}{}^{i}{\!}_{j}\right){R(S)}_{ba}{\!}^j+\thc \nonumber \\
& \;\;\;\;\;\;\;\;\;\;\;\;\;\;\;\;\;\;\;\;\;\;\;\;\;\;\;\;\;\;\;\;\;\;\;\;\;\;\;\;\;\;\;\;\;\;\;\;\;\;\;\;\;\;\;\;\;\;\;\;=\left[\bar{\epsilon}^{i}D_{a}\sum_{k=2}^{n}\left(e^{-1}\LL^{\scriptscriptstyle{(k)}}_{\scriptscriptstyle{\psi}}{}^{a}{}_{i}\right)+\thc \right]+\OO(n+1)\,,
\end{align}
where $\OO(n+1)$ denote variations whose number of supercovariant fields is equal to or greater than $n+1$. We carry on with the variations containing an explicit K-gauge field. They have to satisfy
\begin{align}\label{eq:cancel-k}
f_{a}{\!}^b\bar{\epsilon}^{i}\gamma_{b}\sum_{j=2}^{n}\LL^{\scriptscriptstyle{(k)}}_{\scriptscriptstyle{\phi}}{}^{a}{}_{i}-\bar{\epsilon}^{i}\delta_{K \vert_{ f_a}}\sum_{k=2}^{n}\LL^{\scriptscriptstyle{(k)}}_{\scriptscriptstyle{\psi}}{}^{a}{}_{i}+\thc=\OO(n+1)\,.
\end{align} 
The variations containing an explicit gravitino must satisfy
\begin{align}\label{eq:cancel-psi}
& \tfrac{1}{2}[\delta_Q e]\sum_{k=2}^{n}\left(e^{-1}\LL_{0}^{(k)}\right)-\tfrac{e}{2}\eb^{i}\delta_{Q\vert_{\psi_{a}}}\sum_{k=2}^{n}\left(e^{-1}\LL^{\scriptscriptstyle{(k)}}_{\scriptscriptstyle{\psi}}{}^{a}{}_{i}\right)+\tfrac{e}{2}\bar{\psi}_{a}{\!}^{i}\delta_{Q}\sum_{k=2}^{n}\left(e^{-1}\LL^{\scriptscriptstyle{(k)}}_{\scriptscriptstyle{\psi}}{}^{a}{}_{i}\right) \nonumber \\
& -\tfrac{1}{4}\eb_{i}\gamma_{b}\gamma.T^{ji}\sum_{k=2}^{n-1}\left(\LL^{\scriptscriptstyle{(k)}}_{\scriptscriptstyle{\psi^2}}{}^{ab}{}_{jl}\right)\psi_{a}{\!}^{l}-\tfrac{1}{4}\eb^{i}\gamma_{b}\gamma.T_{ji}\sum_{k=2}^{n-1}\left(\LL^{\scriptscriptstyle{(k)}}_{\scriptscriptstyle{\psi^2}}{}^{ab}{}^{j}{\!}_{l}\right)\psi_{a}{\!}^{l}\nonumber \\
& -\tfrac{1}{4} \eb^{i}\sum_{k=2}^{n-1}\left(\LL^{\scriptscriptstyle{(k)}}_{\scriptscriptstyle{\psi^2}}{}^{ab}{}_{ij}\right)\gamma.T^{jl}\gamma_{b}\psi_{al} -\tfrac{1}{4}\eb_{i}\sum_{k=2}^{n-1}\left(\LL^{\scriptscriptstyle{(k)}}_{\scriptscriptstyle{\psi^2}}{}^{ab}{}^{i}{\!}_{j}\right)\gamma.T^{jl}\gamma_{b}\psi_{al} \nonumber \\
& +\tfrac{1}{2}\bar{\psi}_{bi}\sum_{k=2}^{n-1}\left(\LL^{\scriptscriptstyle{(k)}}_{\scriptscriptstyle{\psi\phi}}{}^{ab}{}^{i}{\!}_{j}\right)\delta^{\scriptscriptstyle{(cov)}}_{Q}\phi_{a}{\!}^{j}+\tfrac{1}{2}\bar{\psi}_{b}{\!}^{i}\sum_{k=2}^{n-1}\left(\LL^{\scriptscriptstyle{(k)}}_{\scriptscriptstyle{\psi\phi}}{}^{ab}{}_{ij}\right)\delta^{\scriptscriptstyle{(cov)}}_{Q}\phi_{a}{\!}^{j} \nonumber \\
& +\tfrac{1}{2}\bar{\epsilon}_{i}\sum_{k=2}^{n-1}\left(\LL^{\scriptscriptstyle{(k)}}_{\scriptscriptstyle{\psi\phi}}{}^{ab}{}^{i}{\!}_{j}\right)\delta^{\scriptscriptstyle{(cov)}}_{Q\vert_{\psi_{a}}}\phi_{b}{\!}^{j}+\tfrac{1}{2}\bar{\epsilon}^{i}\sum_{k=2}^{n-1}\left(e\LL^{\scriptscriptstyle{(k)}}_{\scriptscriptstyle{\psi\phi}}{}^{ab}{}_{ij}\right)\delta^{\scriptscriptstyle{(cov)}}_{Q\vert_{\psi_{a}}}\phi_{b}{\!}^{j}+\thc \nonumber \\
& \;\;\;\;\;\;\;\;= {e}\left[\eb^{i}D_{b}\sum_{k=2}^{n}\left(e^{-1}\LL^{\scriptscriptstyle{(k)}}_{\scriptscriptstyle{\psi^2}}{}^{ab}{}_{ij}\right)\psi_{a}{\!}^{j}+\eb_{i}D_{b}\sum_{k=2}^{n}\left(e^{-1}\LL^{\scriptscriptstyle{(k)}}_{\scriptscriptstyle{\psi^2}}{}^{ab}{}^{i}{\!}_{j}\right)\psi_{a}{\!}^{j}+\thc\right]+\OO(n+1)\,.
\end{align}
We continue with the variations containing a bare S-gauge field
\begin{align}\label{eq:cancel-phi}
& -\tfrac{1}{2}\eb^{i}\delta_{S\vert_{\phi_a}}\sum_{k=2}^{n}\left(e^{-1}\LL^{\scriptscriptstyle{(k)}}_{\scriptscriptstyle{\psi}}{}^{a}{}_{i}\right)+\tfrac{1}{2}\bar{\phi}_{a}{\!}^{i}\delta_{Q}\sum_{k=2}^{n}\left(e^{-1}\LL^{\scriptscriptstyle{(k)}}_{\scriptscriptstyle{\phi}}{}^{a}{}_{i}\right)\nonumber\\
&-\tfrac{1}{2}\eb^{i}\sum_{k=2}^{n}\left(e^{-1}\LL^{\scriptscriptstyle{(k)}}_{\scriptscriptstyle{\psi^2}}{}^{ab}{}_{ij}\right)\gamma_{b}\phi_{a}{\!}^{j}-\tfrac{1}{2}\eb_{i}\sum_{k=2}^{n}\left(e^{-1}\LL^{\scriptscriptstyle{(k)}}_{\scriptscriptstyle{\psi^2}}{}^{ab}{}^{i}{\!}_{j}\right)\gamma_{b}\phi_{a}{\!}^{j}\nonumber \\
&  -\tfrac{1}{12}\bar{\epsilon}^{i}\sum_{k=2}^{n-1}\left(e^{-1}\LL^{\scriptscriptstyle{(k)}}_{\scriptscriptstyle{\psi\phi}}{}^{ab}{}_{ij}\right)\gamma_{b}\gamma.T^{jl}\phi_{al}-\tfrac{1}{12}\bar{\epsilon}_{i}\sum_{k=2}^{n-1}\left(e^{-1}\LL^{\scriptscriptstyle{(k)}}_{\scriptscriptstyle{\psi\phi}}{}^{ab}{}^{i}{\!}_{j}\right)\gamma_{b}\gamma.T^{jl}\phi_{al}\nonumber \\
& -\tfrac{1}{4}\bar{\epsilon}_{j}\gamma_{b}\gamma.T^{ij}\sum_{k=2}^{n-1}\left(e^{-1}\LL^{\scriptscriptstyle{(k)}}_{\scriptscriptstyle{\psi\phi}}{}^{ab}{}_{il}\right)\phi_{a}{\!}^{l} -\tfrac{1}{4}\bar{\epsilon}^{j}\gamma_{b}\gamma.T_{ij}\sum_{k=2}^{n-1}\left(e^{-1}\LL^{\scriptscriptstyle{(k)}}_{\scriptscriptstyle{\psi\phi}}{}^{ab}{}^{i}{\!}_{l}\right)\phi_{a}{\!}^{l}+\thc\nonumber \\
&\;\;\;\;\;\;\;\;\;\;\;\;= \left[\eb^{i}D_{b}\sum_{k=2}^{n}\left(e^{-1}\LL^{\scriptscriptstyle{(k)}}_{\scriptscriptstyle{\psi\phi}}{}^{ab}{}_{ij}\right)\phi_{a}{\!}^{j}+\eb_{i}D_{b}\sum_{k=2}^{n}\left(e^{-1}\LL^{\scriptscriptstyle{(k)}}_{\scriptscriptstyle{\psi\phi}}{}^{ab\,i}{\!}_{j}\right)\phi_{a}{\!}^{j}+\thc\right]+\OO(n+1)\,.
\end{align}

The Lagrangian \eqref{eq:Lagrangian-exact} is build iteratively using the equations \eqref{eq:cancel-supercov}-\eqref{eq:cancel-phi}. The first step of the iterative procedure starts at the lowest-order, i.e. at $n=2$.  At this point, the left-hand side of equation \eqref{eq:cancel-supercov} obviously only contains the first term and the expression of $\LL_0^{\scriptscriptstyle{(2)}}$ is already know as it corresponds to the quadratic Lagrangian given in \eqref{eq:quad-action}. This allows us to derive $\LL^{\scriptscriptstyle{(2)}}_{\scriptscriptstyle{\psi}}{}^{a}{}_{i}$. Subsequently, $\LL^{\scriptscriptstyle{(2)}}_{\scriptscriptstyle{\phi}}{}^a{\!}_i$ and $\LL^{\scriptscriptstyle{(2)}}_{\scriptscriptstyle {\psi^2}}{}^{ab}{\!}_{ij}, \LL^{\scriptscriptstyle{(2)}}_{\scriptscriptstyle {\psi^2}}{}^{ab\,i}{\!}_j$  are determined\footnote{We actually find that $\LL^{\scriptscriptstyle{(2)}}_{\scriptscriptstyle{\phi}}{}^a{\!}_i$ vanishes. This is because $\LL^{\scriptscriptstyle{(2)}}_{\scriptscriptstyle{\psi}}{}^{a}{}_{i}$ turns out to be K-invariant.} by imposing \eqref{eq:cancel-k} and \eqref{eq:cancel-psi}, respectively. This, in turn, allows to compute $\LL^{\scriptscriptstyle{(2)}}_{\scriptscriptstyle {\psi\phi}}{}^{ab}{\!}_{ij}$ and $\LL^{\scriptscriptstyle{(2)}}_{\scriptscriptstyle {\psi\phi}}{}^{ab\,i}{\!}_j$ from \eqref{eq:cancel-phi}.

At the $(n-1)$th iteration step, we consider the cancellation of the supersymmetry variations of order $n$ in the supercovariant fields. We start with equation \eqref{eq:cancel-supercov}, where every term on the left-hand side is known from previous iterations, except for $\LL_{0}^{(n)}$. At this stage, one has to determine $\LL_{0}^{(n)}$ so that the whole left-hand side cancels at order $n$ up to a total supercovariant derivative. The quantity on which the derivative acts upon is then $\LL^{\scriptscriptstyle{(n)}}_{\scriptscriptstyle{\psi}}{}^{a}{\!}_{i}$. This will then lead to $\LL^{\scriptscriptstyle{(n)}}_{\scriptscriptstyle{\phi}}{}^{a}{}_{i}$, $\LL^{\scriptscriptstyle{(n)}}_{\scriptscriptstyle{\psi^2}}{}^{ab}{}_{ij}$, $\LL^{\scriptscriptstyle{(n)}}_{\scriptscriptstyle{\psi^2}}{}^{ab}{}^{i}{\!}_{j}$, $\LL^{\scriptscriptstyle{(n)}}_{\scriptscriptstyle{\psi\phi}}{}^{ab}{}_{ij}$ and $\LL^{\scriptscriptstyle{(n)}}_{\scriptscriptstyle{\psi\phi}}{}^{ab}{}^{i}{\!}_{j}$ by solving the equations \eqref{eq:cancel-k}, \eqref{eq:cancel-psi} and \eqref{eq:cancel-phi}. It is important to mention that at every step of the iteration, the equations \eqref{eq:cancel-supercov}--\eqref{eq:cancel-phi} should be solved one after the other as each equation requires an input obtained by solving the previous one. In this way, we build all the terms of the Lagrangian \eqref{eq:Lagrangian-exact} up to quadratic order in the fermion fields.

\setcounter{equation}{0}

\section{Results and discussion}\label{sec:results}
In this section, we present all the supercovariant terms of the $N=4$ conformal supergravity Lagrangian up to quadratic order in the fermion fields, obtained through the iterative procedure presented in section \ref{sec:strategy}. For the reader's convenience, the interactions involving explicit gauge fields are given in appendix \ref{App:Add-results}. 

In section \ref{sec:strategy}, we argued the Lagrangian takes the form \eqref{eq:Lagrangian-structure}. Within this scheme, the purely supercovariant terms at all order in the fields, bosonic or fermionic, are cast within the quantity denoted by $\mathcal{L}_0$. Let us now split $\mathcal{L}_0$ into
\begin{equation}\label{L-cov}
\LL_0=\LL_{\text{Q}}+\LL_{\text{B}}+\LL_{\text{F}}+\ldots\,,
\end{equation}
where $\LL_{\text{Q}}$, $\LL_{\text{B}}$ and $\LL_{\text{F}}$ are respectively the quadratic Lagrangian \eqref{eq:quad-action}, all the purely bosonic supercovariant interaction terms and the supercovariant interaction terms quadratic in the fermion fields. Here, the dots denote terms which are quartic, sextic and octic in the fermion fields and which, therefore, are outside of the scope of this paper.

We first recall the quadratic Lagrangian
\begin{align}\label{L-quad}
  e^{-1}\mathcal{L}_0 =&\,\tfrac{1}{2}{R}(M)^{abcd}\,R(M)^{-}{\!\!\!\!}_{abcd} +
  {R}(V)^{ab \,i}{}_j \, {R}(V)_{ab}{\!\!\!\!\!}^{-\,\, j} {}_i \nonumber \\[1mm]
  &\, -4\, T_{ab}{\!}^{ij}\,D^{a}D_{c}T^{cb}{\!}_{ij}+\tfrac{1}{4}
  E_{ij}\,D^2E^{ij} +\tfrac18D_{ij}{\!}^{kl}\,D_{kl}{}^{ij} \nonumber \\[1mm]
  &\,
  -2\,\bar P^{a} \big[D_{a}D^{b} P_{b}+D^{2} P_{a}\big]-2\,
  D^{a}P^{b} D_{a}\bar  P_{b} - D^{a}P_{a} D^{b}\bar
  P_{b} \nonumber\\[1mm]
  &\,+\bar{R}(Q)_{ab}{}^{i}R(S)^{ab}{}_{i}-\bar{\chi}^{ij}{}_{k}\Dslash\chi^{k}_{~ij}-\tfrac{1}{2}\bar{\Lambda}_{i}\left(D^{2}\Dslash+\Dslash D^{2}-\Dslash^{3}\right)\Lambda^{i}\,+\thc\,,
\end{align}
which was discussed in section \ref{sec:quadratic-lagrangian} and served as the basis for the iterative procedure. 

The bosonic interaction terms at all order in the fields are 
\begin{align}\label{L-b}
 e^{-1} \mathcal{L}_{\text{B}} =& \,
     \tfrac13 P^{a}\,\bar P_{a} \, P^b \,\bar P_{b}
    +  P^{a}\, P_{a}\, \bar P^{b}\, \bar P_{b} \nonumber \\[1mm]
    &\,
    -\tfrac{1}{16} E_{ij}\,E^{jk}\,E_{kl}\, E^{li}
    +\tfrac{1}{48}\big[E_{ij}\,E^{ij}\big]^2  \nonumber \\[1mm]
    & \,-\tfrac16 E_{ij}\,E^{ij}\,P^{a}\,\bar P_{a}
    -8\,T^{ab\,ij}\, T_{bc\,ij}\, P_{a}\, \bar P^{c} \nonumber \\[1mm]
   &\,+T^{ab\,ij}\, T_{ab}{\!}^{kl} \, T_{cd\,ij}\, T^{cd}{\!}_{kl}
    -T^{ab\,ik}\, T_{ab}{\!}^{jl}\, T_{cd\,ij}\, T^{cd}{\!}_{kl} \nonumber \\[1mm]
    &\,+\varepsilon^{ijkl}\,T^{ab}{\!}_{ij}\,E_{km} \, R(V)_{ab}{\!}^m{}_l  \nonumber \\[1mm]
    &\, -\varepsilon^{ijkl}  \bar P^c \, [4\, D_a T^{ab}{\!}_{ij}
      \, T_{bc\,kl}
    - D_c T^{ab}{\!}_{ij} \, T_{ab\,kl} ] \nonumber \\[1mm]
    &\,-\tfrac18\varepsilon_{ijkl}\,\varepsilon_{mnpq}\,T^{ab\,ij}\,
    T_{ab}{\!}^{mn}\, E^{kp}\, E^{lq}
    +\mathrm{h.c.}\,,
    \end{align}
which involve cubic and quartic terms in the fields. Quintic terms are forbidden due to the Weyl weights of the bosons.

The interaction terms which are quadratic in the fermion fields read
\begin{align}\label{L-f}
e^{-1}\LL_{\text{F}} =  &\,\, \tfrac14\varepsilon_{ijkl}\bar{\chi}^{ij}{\!}_{m}\gamma\cdot T^{kl}\Dslash\Lambda^{m}-\tfrac14\varepsilon_{ijkl}\bar{\chi}^{ij}{\!}_{m}\gamma\cdot T^{kl}\overleftarrow{\Dslash}\Lambda^{m} \nonumber\\[1mm]
&-\tfrac{1}{2}\varepsilon_{ijkl}\bar{\chi}^{ij}{\!}_{m}\chi^{kl}{\!}_{n}E^{mn}-\tfrac34\varepsilon^{ijkl}\bar{{R}}(Q)_{i}\cdot \Dslash T_{jk}\Lambda_{l}-\tfrac32\varepsilon^{ijkl}\bar{{R}}(Q)_{i}\overleftarrow{\Dslash}\cdot T_{jk}\Lambda_{l} \nonumber\\[1mm]
&+2\,D^{a}\bar{\Lambda}^{i}{R}(Q)_{abi}\bar{P}^b+\tfrac12\bar{\Lambda}_{j}\gamma^{b}\Lambda^{i}D^{a}{R}(V)_{ab}{}^{j}{\!}_{i}+\varepsilon^{ijkl}\bar{\chi}^{mn}{\!}_{l}\Lambda_{i}T_{mn}\cdot T_{jk}\nonumber\\[1mm]
& +\bar{\chi}^{ij}{\!}_{k}\gamma^{a}\gamma\cdot T_{ij}\Lambda^{k}\bar{P}_{a}+ \tfrac16\varepsilon_{ijkl}\bar{\chi}^{ij}{\!}_{m}\Lambda_{n}E^{mk}E^{ln}-\tfrac13\varepsilon_{ijkl}\bar{\chi}^{ij}{\!}_{m}\gamma^{a}\Lambda^{k}E^{ml}\bar{P}_{a}\nonumber\\[1mm]
&-\tfrac{1}{12} \bar{\Lambda}_{i}\Dslash\Lambda^{i}E_{jk}E^{jk}+\tfrac13\bar{\Lambda}_{i}\Dslash\Lambda^{j}E_{jk}E^{ki}-\tfrac16\bar{\Lambda}_{i}\gamma^{a}\Lambda^{j}D_{a}E_{jk}E^{ki}+\tfrac56\bar{\Lambda}_{i}\Lambda_{j}D_{a}E^{ij}P^{a}\nonumber\\[1mm]
&+\tfrac23\bar{\Lambda}_{i}\Lambda_{j}E^{ij}D_{a}P^{a}+\tfrac13\bar{\Lambda}_{i}\gamma^{ab}D_{a}\Lambda_{j}E^{ij} P_{b}+\tfrac43\bar{\Lambda}_{i}\gamma_{a}D_{b}\Lambda^{i}\bar{P}^{b}P^{a}-\tfrac16\bar{\Lambda}_{i}\gamma_{a}\Lambda^{i}D_{b}\bar{P}^{a} P^{b} \nonumber\\[1mm]
&-\tfrac16\bar{\Lambda}_{i}\gamma_{a}\Lambda^{i}D_{b}\bar{P}^{b} P^{a}+\tfrac23\bar{\Lambda}_{i}\Dslash\Lambda^{i}P_{b}\bar P^{b}+\tfrac43\varepsilon^{abcd}\bar{\Lambda}_{i}\gamma_{a}D_{b}\Lambda^{i}\bar{P}_{c}P_{d}-2\,D_a \bar\Lambda_i\ga^c\Lambda^i T^{ab}{\!}_{jk}T_{cb}{\!}^{jk}\nonumber\\[1mm]
&-2\, \bar\Lambda_i\ga^c\Lambda^i D_a T^{ab}{\!}_{jk}T_{cb}{\!}^{jk}+2\,D_a\bar\Lambda_i\ga^c\Lambda^j T^{ab}{\!}_{jk}T_{cb}{\!}^{ik}+2\,\bar\Lambda_i\ga^c\Lambda^j D_a T^{ab}{\!}_{jk}T_{cb}{\!}^{ik}\nonumber\\[1mm]
&-\tfrac23\varepsilon^{ijkl}\bar{\Lambda}_{i}D^{a}\Lambda_{j}T_{abkl}P^{b}+\varepsilon^{ijkl}\bar{\Lambda}_{i}\gamma^{ab}\Lambda_{j}D_{a}T_{bckl}P^{c}+\tfrac23\varepsilon^{ijkl}\bar{\Lambda}_{i}\gamma^{ab}\Lambda_{j}T_{bckl}D_{a}P^{c}\nonumber\\[1mm]
&-\tfrac23\varepsilon^{ijkl} D^a \bar\Lambda_i\ga^b\Lambda^m E_{mj}T_{abkl}+\varepsilon_{ijkl}\bar\Lambda_m\ga^b\Lambda^i D^a E^{jm}T_{ab}{}^{kl}+\tfrac13\varepsilon_{ijkl}\bar\Lambda_m\ga^b\Lambda^i  E^{jm}D^a T_{ab}{\!}^{kl}\nonumber\\[1mm]
&-\tfrac18\varepsilon^{klmn}\bar\Lambda_i\Lambda_jT_{kl}\cdot T_{mn}E^{ij}+\tfrac16 \varepsilon^{klmn}\bar\Lambda_k\Lambda_i T_{jl}\cdot T_{mn}E^{ij}+\tfrac23\bar\Lambda_i\ga^a\Lambda^jE_{jk}T_{ab}{}^{ki}{P}^{b} \nonumber \\[1mm]
& -\tfrac13\bar\Lambda_i\ga^{ab}\Lambda_j T_{ab}{\!}^{ij}{P}^{c}{P}_{c} - \tfrac{1}{12}\varepsilon_{jklm}\bar\Lambda_i\ga^a\Lambda^i T^{jk}\cdot T^{lm}{P}_{a}+\thc\,.
\end{align}
They involve cubic, quartic and quintic terms in the fields. Note that there are no terms of sextic, septic or octic order in the fields as, due to the restrictions on the Weyl weights, these would be of higher-order in the fermion fields. Finally, \eqref{L-quad}, \eqref{L-b} and \eqref{L-f} are $\mathrm{SU}(1,1)$ invariant and their sum is K-invariant.

As we already mentioned in section \ref{sec:intro}, the bosonic part of the $N=4$ conformal supergravity Lagrangian has been derived in \cite{Buchbinder:2012uh}.  Because it was obtained in a different set of conventions, we have converted their result in the conventions of the present paper to facilitate the comparison with our results. In particular, this requires to covariantize the curvatures and derivatives with respect to the conformal boosts and to switch to a different parametrisation of the coset space. Up to a Gauss-Bonnet term, the Lagrangian in \cite{Buchbinder:2012uh} is then equivalent to
\begin{align}\label{eq:Tseytlin}
e^{-1}\LL =&\,\tfrac{1}{2}{R}(M)^{abcd}\,R(M)^{-}{\!\!\!\!}_{abcd} +
  {R}(V)^{ab \,i}{}_j \, {R}(V)_{ab}{\!\!\!\!\!}^{-\,\, j} {}_i \nonumber \\[1mm]
  &\, -4\, T_{ab}{\!}^{ij}\,D^{a}D_{c}T^{cb}{\!}_{ij}+\tfrac{1}{4}
  E_{ij}\,D^2E^{ij} +\tfrac18D_{ij}{\!}^{kl}\,D_{kl}{}^{ij}  \nonumber \\[1mm]
  &\,
  -2\,\bar P^{a} \big[D_{a}D^{b} P_{b}+D^{2} P_{a}\big]-2\,
 D^{a}P^{b} D_{a}\bar  P_{b} - D^{a}P_{a}D^{b}\bar
  P_{b}  \nonumber\\
  &\,+\tfrac43 P^{a}\,\bar P_{a} \, P^{b}\,\bar P_{b}+P^{a}\,P_{a}\,\bar P^{b}\,\bar
      P_{b} \nonumber \\[1mm]
      &\,-\tfrac{1}{24}E_{ij}\, E^{jk}\,E_{kl}\,E^{li} \nonumber \\[1mm]
      &\,+\tfrac{1}{12} E_{ij}E^{ij}\, P^{a}\bar P_{a}
      -4\,T^{ab\, ij}\,T_{bc\,ij} \,P_{a}\, \bar P^c  \nonumber  \\[1mm]
      & \,+\tfrac{5}{12}T_{ab}{\!}^{ij}\, T^{ab\,kl}\, T^{cd}{\!}_{ij}\,
      T^{cd\,kl} +\tfrac{1}{6}T^{ab\,ik}\, T_{ab}{\!}^{jl}\,
      T_{cd\,ij}\, T^{cd}{\!}_{kl}+\mathrm{h.c.}\,.
\end{align}
We now compare the above expression with \eqref{L-b} and the bosonic part of \eqref{L-quad}. 

Clearly, the quadratic Lagrangians agree as the first three lines of \eqref{eq:Tseytlin} coincide with the bosonic part of \eqref{L-quad}. We note, however, a number of differences when comparing interaction terms. The most obvious one is perhaps the presence of terms cubic in the fields in our results while none appear in \eqref{eq:Tseytlin}. Further differences concern the quartic terms in the fields. Indeed, the last term of the second line and the last line in \eqref{L-b} are not present in \eqref{eq:Tseytlin}. Moreover, none of the coefficients of the remaining terms match.

When truncated to $N=2$, the result of \cite{Buchbinder:2012uh} is consistent with the known non-linear Lagrangian of $N=2$ conformal supergravity \cite{Bergshoeff:1980is}. As it turns out, we find that our results also yield the correct $N=2$ Lagrangian upon truncation. However, one must note that for the bosonic action, most of the fields simply disappear in the truncation process. Indeed, there are no $N=2$ descendants of the fields $P_\mu$ and $E^{ij}$. For this reason, the only comparison at the $N=2$ level that can be made of the bosonic sectors concerns the relative coefficient between the kinetic term and the quartic interactions of the field $T_{ab}{\!}^{ij}$. It is surprising that while both results agree at the $N=2$ level, such striking differences are present in the full $N=4$ setting.

As should be clear from the iterative procedure that was used in this paper, the consistency of each term in our result relies on the consistency of many other terms. Therefore, our computation passes a multitude of crosschecks. It should also be noted that all the terms in our result correspond to possible Feynman diagrams of the gauge theory \cite{deRoo:1984gd} with logarithmically divergent contributions. 
\newline

\underline{Note added}: After submitting this paper to the arXiv, it was found that several terms were missed in the last stages of the computation of \cite{Buchbinder:2012uh}. The authors of \cite{Buchbinder:2012uh}, in particular A. Tseytlin, were kind enough to confirm this observation. Once repaired, these omissions precisely match with the corresponding terms in \eqref{L-b}.

\section*{Acknowledgements}
We would like to thank Bernard de Wit and Daniel Butter for valuable discussions and careful reading of the manuscript. We are also grateful to A. Tseytlin for a discussion of the results.  F. C. thanks IISER Thiruvananthapuram and Prof. Shankaranarayanan for the hospitality extended to him during the completion of this work. B. S. thanks Nikhef Amsterdam for the hospitality extended to him during the course of this work. This work is supported by the ERC Advanced
Grant no. 246974, ``{\it{Supersymmetry: a window to non-perturbative physics}} ''.


\begin{appendix}

\setcounter{equation}{0}

\section{Terms with explicit fermionic gauge fields}
\label{App:Add-results}
In this section, we present all the terms at quadratic order in the fermion fields which contain explicit fermionic gauge fields. Therefore, we give the expression for the supercovariant quantities 
\begin{equation}
\LL_\psi{\!}^a{\!}_i\,,\;\LL_\phi{\!}^a{\!}_i\,,\;\LL_{\psi^2}{\!}^{ab}{\!}_{ij}\,,\;\LL_{\psi^2}{\!}^{ab\,i}{\!}_j\,,\;\LL_{\psi\phi}{\!}^{ab}{\!}_{ij}\,,\;\LL_{\psi\phi}{\!}^{ab\,i}{\!}_j\,,
\end{equation}
which, as described in \eqref{eq:Lagrangian-exact}, appear in the Lagrangian coupled to bare fermionic gauge fields. For the purpose of this paper, we can restrict ourselves to the terms in $\LL_\psi{\!}^a{\!}_i$ and $\LL_\phi{\!}^a{\!}_i$ which are linear in fermions. Likewise, it is enough to only consider the bosonic terms in $\LL_{\psi^2}{\!}^{ab}{\!}_{ij},\LL_{\psi^2}{\!}^{ab\,i}{\!}_j,\LL_{\psi\phi}{\!}^{ab}{\!}_{ij},\LL_{\psi\phi}{\!}^{ab\,i}{\!}_j\,$.

Let us first consider $\LL_\psi{\!}^a{\!}_i$ which is contracted with a gravitino in the Lagrangian. For the reader's convenience, we split this quantity into
\begin{equation}\label{M}
\LL_\psi{\!}^a{\!}_i=\,\LL^{\text{\tiny{(2)}}}_\psi{}^a{\!}_i+\LL^{\text{\tiny{(3)}}}_\psi{}^a{\!}_i+\LL^{\text{\tiny{(4)}}}_\psi{}^a{\!}_i +\ldots\,,
\end{equation}
where $\LL^{\text{\tiny{(2)}}}_\psi{}^a{\!}_i,\LL^{\text{\tiny{(3)}}}_\psi{}^a{\!}_i$ and $\LL^{\text{\tiny{(4)}}}_\psi{}^a{\!}_i $ are quadratic, cubic and quartic in the fields, respectively. Due to Weyl weight restrictions, the dots denote terms which are of higher-order in fermions. The quadratic part reads
\begin{align}\label{M-quad}
e^{-1}\LL^{\text{\tiny{(2)}}}_\psi{}^a{\!}_i= & \, \ga^{a}\chi^{l}{\!}_{jk}D^{jk}{\!}_{li}+\tfrac12\ga^{a}\gamma\cdot T_{jk}\Dslash\chi^{jk}{\!}_{i}+2\ga^{a}R(Q)_{cdj}R(V)^{cdj}{\!}_{i}+\tfrac12\ga^{a}\gamma_{eb}R(Q)_{cdi}R(M)^{ebcd}  \nonumber \\[1mm]
& -2\ga^{a}R(S)_{cd}{}^{j}T^{cd}{\!}_{ij}-\epsilon_{ijkl}\ga^{a}\ga_{bd}\Lambda^{j}D^{b}D_{c}T^{cdkl}+\tfrac{1}{2}\epsilon_{ijkl}\ga^{a}\Dslash\chi^{jk}{\!}_{m}E^{lm}+\tfrac{1}{2}\ga^{a}\Lambda^{j}D^{2}E_{ij}\nonumber \\[1mm]
&+\ga^{a}\big[\left(D_{d}\Dslash+\Dslash D_{d}+\ga_{d}D^{2}\right)\Lambda_{i}\big]{P}^{d}+\tfrac{1}{2}\ga^{a}\ga^{d}\Lambda_{i}\big[D_{d}D_{b}{P}^{b}+D^{2}{P}_{d}\big]+\ga^{a}\Dslash \Lambda_{i}D_{d}{P}^{d}\nonumber \\[1mm]
& +2\ga^{a}\ga^{d}D^{b}\Lambda_{i}D_{(d}{P}_{b)}+\ga^{a}\gamma\cdot R(V)^{j}{\!}_{k}\chi^{k}{\!}_{ij}\,,
\end{align}
while the cubic part is 
\begin{align}\label{M-cub}
e^{-1}\LL^{\text{\tiny{(3)}}}_\psi{}^a{\!}_i= & \, -\tfrac{1}{2}\ga^{a}\chi^k{\!}_{ji}E_{kl}E^{lj}-\gamma_{b}{R}(Q)_{i}\cdot T_{jk}T^{ab\,jk}-6\,\gamma_{b}{R}(Q)_{k}\cdot T_{ij}T^{ab\,jk}+4\,\gamma^{b}\chi^{l}{\!}_{jk}T_{db\,li}T^{ad\,jk}\nonumber \\[1mm]
& -\varepsilon_{jklm}\gamma^{c}\gamma_{d} \chi_{i}{\!}^{jk}{P}_{c}T^{ad\,lm}+\varepsilon_{jklm}\gamma\cdot T^{lm}\chi^{jk}{\!}_i{P}^{a} -2\,\varepsilon_{ijkl}\gamma^{c}\gamma^{a}{R}(Q)^{l}\cdot T^{jk}{P}_{c}\nonumber \\
& +3\,\varepsilon_{ijkl}{R}(Q)^{l}\cdot T^{jk}{P}^{a}+\tfrac{1}{2}\varepsilon^{jklm}\gamma_{b}\chi^{n}{\!}_{jk}E_{ni}T^{ab}{\!}_{lm}-\varepsilon^{jklm}\gamma_{b}\chi^{n}{\!}_{ij}E_{kn}T^{ab}{\!}_{lm}\nonumber \\[1mm]
& -\varepsilon_{jklm}\gamma_{b}\chi^{j}{\!}_{ni}E^{kn}T^{ab\,lm}-\tfrac{1}{12}\gamma^{a}\gamma\cdot {R}(V)^{j}{\!}_{i}\Lambda^{k}E_{jk}-\tfrac{1}{6}\gamma.\gamma^{a}{R}(V)^{j}{\!}_{i}\Lambda^{k}E_{jk}\nonumber \\[1mm]
& +\tfrac{1}{4}\gamma^{a}\gamma\cdot {R}(V)^{j}{\!}_{k}\Lambda^{k}E_{ji}+\tfrac{1}{4}\varepsilon^{jklm}\gamma^{a}{R}(Q)_{j}\cdot T_{kl}E_{mi}+2{R}(Q)^{ab\,j}{P}_{b}E_{ij}\nonumber \\[1mm]
& +\gamma_{cd}\Lambda_{i}{R}(M)^{abcd}{P}_{b}+\gamma^{c}{R}(Q)^{ab}{\!}_i\bar{P}_{c}{P}_{b} -\tfrac{19}{3}\varepsilon^{jklm}\Lambda_{m}D^b T_{bc\,li}T^{ac}{\!}_{jk} \nonumber \\[1mm]
& -6\,\varepsilon^{jklm}D^b \Lambda_m T_{bc\,li}T^{ac}{\!}_{jk}+3\,\varepsilon^{jklm}\Lambda_{m}D^b T^{ac}{\!}_{li}T_{bc\,jk}+2\,\varepsilon^{jklm}\Lambda_{m} T^{ac}{\!}_{li}D^bT_{bc\,jk} \nonumber\\[1mm]
& -2\,\varepsilon^{jklm}\Lambda_{m} D^a \big[T_{li}\cdot T_{jk}\big]-2\, \varepsilon^{jklm}\gamma_{dc}D_b \Lambda_{m}T^{bd}{\!}_{li}T^{ac}{\!}_{jk} +\tfrac{5}{3}\varepsilon^{jklm}\gamma_{dc} \Lambda_{m} D_b T^{bd}{\!}_{li}T^{ac}{\!}_{jk} \nonumber \\[1mm]
& -2\,\varepsilon^{jklm}\gamma_{dc}\Lambda_{m} T^{ad}{\!}_{li}D_{b}T^{bc}{\!}_{jk} +\varepsilon^{jklm}\gamma_{dc} \Lambda_{m} D_b T^{ad}{\!}_{li}T^{bc}{\!}_{jk}+\tfrac12 \varepsilon_{ijkl}\ga_b\Lambda^{l} T_{cd}{\!}^{jk}{R}(M)^{abcd} \nonumber \\[1mm]
& + 2\,\Lambda_{k}D_{c}E^{jk}T^{ca}{\!}_{ji}+2\,\Lambda_{k}E^{jk}D_{c}T^{ca}{\!}_{ji} +\tfrac{1}{3}\gamma^{ba}\Lambda_{k}D^{c}E^{jk}T_{cb\,ji} +\tfrac{1}{3}\gamma^{ba}D^{c}\Lambda_{k}E^{jk}T_{cb\,ji}\nonumber \\[1mm]
& -\tfrac{1}{3}\gamma^{ba}\Lambda_{k}E^{jk}D^{c}T_{cb\,ji} +\tfrac{16}{3}\gamma^{c}\Lambda^{j}\bar{P}^{b}D^{a}T_{bcji}+\tfrac{16}{3}\gamma^{c}\Lambda^{j}D^{a}\bar{P}^{b}T_{bcji}+\tfrac{4}{3}\gamma^{c}D^{a}\Lambda^{j}\bar{P}^{b}T_{bcji} \nonumber \\[1mm]
& +\tfrac{2}{3}\gamma_{c}D_{b}\Lambda^{j}\bar{P}^{b}T^{ac}{\!}_{ji}+\tfrac{10}{3}\gamma^{c}\Lambda^{j}\bar{P}^{a}D^{b}T_{bcji}-\tfrac{10}{3}\gamma_{c}\Lambda^{j}\bar{P}^{b}D_{b}T^{ac}{\!}_{ji}+\tfrac{2}{3}\gamma^{c}D^{b}\Lambda^{j}\bar{P}^{a}T_{bcji} \nonumber \\[1mm]
&+\tfrac{2}{3}\gamma_{c}\Lambda^{j}D^{b}\bar{P}_{b}T^{ac}{\!}_{ji}+\tfrac{2}{3}\gamma^{c}\Lambda^{j}D^{b}\bar{P}^{a}T_{bcji} +\tfrac{4}{3}\gamma^{b}\Lambda^{j}\bar{P}_{b}D_{c}T^{ca}{\!}_{ji}-\tfrac{20}{3}\gamma^{a}\Lambda^{j}\bar{P}^{b}D^{c}T_{cbji}\nonumber \\[1mm]
& +\tfrac{28}{3}\gamma^{b}\Lambda^{j}\bar{P}_{c}D_{b}T^{ac}{\!}_{ji}-\tfrac{8}{3}\gamma^{a}D^{c}\Lambda^{j}\bar{P}^{b}T_{cbji}+\tfrac{16}{3}\gamma^{b}\Lambda^{j}D_{b}\bar{P}_{c}T^{ac}{\!}_{ji}+\tfrac{4}{3}\gamma^{b}D_{b}\Lambda^{j}\bar{P}_{c}T^{ac}{\!}_{ji} \nonumber \\[1mm]
& +\tfrac14\big[\tfrac13\ga_{cd}\ga^{ab}+\ga^{ab}\ga_{cd}\big]\Lambda_j R (V)^{cd}{}^j{\!}_{i}{P}_{b}-\tfrac16\ga_{cd}\ga^{ab}\Lambda_j R (V)^{cd}{}^j{\!}_{i}{P}_{b}-\tfrac12\ga_{cd}\Lambda_{j}R (V)^{cd}{}^j{\!}_{i}{P}^{a}\nonumber \\[1mm]
& -\tfrac13\Lambda_{j}R (V)^{ba\,j}{\!}_{i}{P}_{b}-\tfrac13\ga_{cd}\Lambda_{j}R (V)^{ca\,j}{\!}_{i}{P}^{d}-4\,\epsilon_{ijkl}\ga^c\Lambda^{j}T_{cb}{\!}^{km}R(V)^{ab\,l}{\!}_{m}\nonumber\\[1mm]
&-\tfrac12\epsilon_{ijkl}\ga^a\Lambda^mT^{ef}{}^{jk}R(V)_{ef}{}^{l}{\!}_{m}+\tfrac23\epsilon_{jklm}\ga_b\Lambda^{m}T_{c}{}^{[a\,kl}R(V)^{b]c\,j}{\!}_{i}\nonumber\\[1mm]
&\,+\tfrac56\epsilon_{jklm}\ga^a\Lambda^{m}T^{cd\,kl}R(V)_{cd}{}^{j}{\!}_{i}\,.
\end{align}
The quartic part takes the following form
\begin{align}\label{M-quart}
e^{-1}\LL^{\text{\tiny{(4)}}}_\psi{}^a{\!}_i= & \, \tfrac{1}{12}\gamma^{a}\Lambda^{j}E_{ij}E_{kl}E^{kl}-\tfrac{1}{6}\gamma^{a}\Lambda^{k}E_{lk}E_{ij}E^{lj}+\tfrac{1}{6}\Lambda_{i}E_{jk}E^{jk}{P}^{a}+\tfrac{1}{6}\ga^{ab}\Lambda_{i}E_{jk}E^{jk}{P}_{b}\nonumber \\[1mm]
& +\tfrac{2}{3}\Lambda_{k}E_{ij}E^{kj}{P}^{a}-\tfrac{1}{3}\ga^{ab}\Lambda_{k}E_{ij}E^{kj}{P}_{b}-\tfrac{2}{3}\gamma^{a}\Lambda^{j}E_{ij}\bar{P}_{b}P^{b}-\tfrac{4}{3}\gamma^{b}\Lambda^{j}E_{ij}{P}^{a}\bar{P}_{b}\nonumber \\[1mm]
& +\tfrac{10}{3}\gamma^{b}\Lambda^{j}E_{ij}{P}_{b}\bar{P}^{a} -\tfrac{1}{3}\varepsilon^{abcd}\gamma_{d}\Lambda^{j}{P}_{b}\bar{P}_{c}E_{ji}+\tfrac{7}{3}\ga^{bc}\Lambda_{i}{P}^{a}{P}_{b}\bar{P}_{c}-\tfrac{2}{3}\ga^{ac}\Lambda_{i}{P}_{b}{P}_{c}\bar{P}^{b} \nonumber \\[1mm]
& -\tfrac{2}{3}\ga^{ac}\Lambda_{i}{P}_{b}{P}^{b}\bar{P}_{c}-\tfrac{5}{3}\Lambda_{i}{P}_{b}{P}^{b}\bar{P}^{a}+\tfrac{1}{3}\Lambda_{i}{P}_{b}{P}^{a}\bar{P}^{b}-\tfrac{7}{3}\varepsilon_{ijkl}\gamma^{a}\Lambda^{j}T_{bc}{\!}^{kl}{P}^{b}\bar{P}^{c}\nonumber \\[1mm]
& +\tfrac{3}{2}\varepsilon_{ijkl}\gamma_{b}\Lambda^{j}T^{ac}{}^{kl}{P}^{b}\bar{P}_{c} +\tfrac{23}{6}\varepsilon_{ijkl}\gamma_{b}\Lambda^{j}T^{ac}{}^{kl}{P}_{c}\bar{P}^{b}-\tfrac{5}{6}\varepsilon_{ijkl}\gamma^{b}\Lambda^{j}T_{bc}{\!}^{kl}{P}^{a}\bar{P}^{c} \nonumber \\[1mm]
& -\tfrac{1}{2}\varepsilon_{ijkl}\gamma^{b}\Lambda^{j}T_{bc}{\!}^{kl}{P}^{c}\bar{P}^{a}+4\,\varepsilon_{ijkl}\gamma_{b}\Lambda^{j}T^{ba}{}^{kl}\bar{P}_{c}P^{c}-\tfrac16\varepsilon_{ijkl}\ga^b\Lambda^m T_{ba}{\!}^{jk}E_{mn}E^{nl}\nonumber \\[1mm]
&+\tfrac23\varepsilon^{jklm}\ga_b\Lambda^n E_{mn}E_{il}T^{ba}{\!}_{jk}-\tfrac12\varepsilon_{jklm}\ga_b\Lambda^lE_{ni}E^{nm}T^{ba}{}^{jk}+2\,\ga^c\Lambda^l E_{li}T^{ab}{\!}_{jk}T_{cb}{\!}^{jk}\nonumber\\[1mm]
&+2\,\varepsilon_{ijkl}\varepsilon_{mnpq}\ga^c\Lambda^mE^{ln}T^{ab}{}^{pq}T_{cb}{\!}^{jk}+\tfrac12 \varepsilon_{ijkl}\varepsilon_{mnpq}\ga^a\Lambda^n E^{ml}T^{pq} \cdot T^{jk}\nonumber\\[1mm]
& -\tfrac{23}{3}\ga^c\Lambda^lE_{lj}T^{ab}{\!}_{ik}T_{cb}{\!}^{jk}+2\,\ga^c\Lambda^jE_{ik}T^{ab}{\!}_{lj}T_{cb}{\!}^{lk}+\tfrac43\varepsilon^{jklm}\Lambda_l E_{im}T^{ab}{\!}_{jk}{P}_b\nonumber\\[1mm]
&-\tfrac23\varepsilon_{ijkl}\ga_{cb}\Lambda_m E^{ml}T^{ab}{}^{jk}{P}^{c}-2\,\varepsilon^{jklm}\ga_{cb}\Lambda_m E_{il}T^{ab}{\!}_{jk}{P}^c+\tfrac13\varepsilon_{ijkl}\ga^{bc}\Lambda_m E^{ml}T_{bc}{\!}^{jk}{P}^a\nonumber\\[1mm]
&  +\tfrac{46}{3}\Lambda_{k}T^{ab}{}^{kj}T_{bc}{}_{ji}{P}^{c}+\tfrac{28}{3}\Lambda_{i}T^{ab}{}^{jk}T_{bc}{}_{jk}{P}^{c} +\tfrac{34}{3}\gamma^{cd}\Lambda_{k}T^{ab}{}^{kj}T_{bc}{}_{ji}{P}_{d}\nonumber \\[1mm]
&+\tfrac{4}{3}\gamma^{cd}\Lambda_{i}T^{ab}{}^{jk}T_{bc}{}_{jk}{P}_{d}  +\tfrac{14}{3}\ga\cdot T^{kj}\Lambda_{k}T^{ab}{\!}_{ji}{P}_{b} -\tfrac{2}{3}\ga\cdot T^{jk}\Lambda_{i}T^{ab}{\!}_{jk}{P}_{b}\nonumber \\[1mm]
&-2\,\varepsilon_{jklm}\gamma_{b}\Lambda^{l}T^{jk}\cdot T^{mn}T^{ab}{\!}_{in} +\varepsilon_{ijkl}\gamma_{b}\Lambda^{l}T^{jk}\cdot T^{mn}T^{ab}{\!}_{mn}\nonumber\\[1mm]
&+8\,\varepsilon_{jklm}\gamma_{d}\Lambda^{l}T^{ab}{\!}_{in}T_{bc}{\!}^{jk}T^{cd}{}^{mn}\,.
\end{align}

We now present the expression of $\LL_\phi{\!}^a{\!}_i$ which appears contracted with an S-gauge field in the Lagrangian. The terms linear in fermions can only be of cubic order in the fields 
\begin{align}
e^{-1}\LL_\phi{\!}^a{\!}_i= &\, \tfrac12\ga_b\Lambda_j E^{lj}T^{ab}{\!}_{li}+\tfrac{19}{3}\Lambda^j \bar{P}_bT^{ba}{\!}_{ji}-\tfrac{1}{6}\ga^{bc}\Lambda^j \bar{P}^{a}T_{bc\,ji}\nonumber\\[1mm]
&+\tfrac13\ga^{ab}\Lambda^j \bar{P}^c T_{bc}{}_{ji}+\tfrac23\varepsilon^{jklm}\ga^b\Lambda_mT_{bc}{}_{li}T^{ac}{\!}_{jk}+2\,\varepsilon^{jklm}\ga^a\Lambda_m T_{li}\cdot T_{jk}\,.
\end{align}

We move on to $\LL_{\psi^2}{\!}^{ab}{\!}_{ij}$ and $\LL_{\psi^2}{\!}^{ab\,i}{\!}_j$ which enter the Lagrangian contracted with two gravitini. For clarity, we split them into
\begin{align}
\LL_{\psi^2}{\!}^{ab}{\!}_{ij}&=\,\LL^{\text{\tiny{(2)}}}_{\psi^2}{}^{ab}{\!}_{ij}+\LL^{\text{\tiny{(3)}}}_{\psi^2}{}^{ab}{\!}_{ij} +\ldots\,, \\
\LL_{\psi^2}{\!}^{ab\,i}{\!}_j &= \,\LL^{\text{\tiny{(2)}}}_{\psi^2}{}^{ab\,i}{\!}_j+\LL^{\text{\tiny{(3)}}}_{\psi^2}{}^{ab\,i}{\!}_j+\ldots\,,
\end{align}
where $\LL^{\text{\tiny{(2)}}}_{\psi^2}{}^{ab}{\!}_{ij}, \LL^{\text{\tiny{(2)}}}_{\psi^2}{}^{ab\,i}{\!}_j$ and $\LL^{\text{\tiny{(3)}}}_{\psi^2}{}^{ab}{\!}_{ij},\LL^{\text{\tiny{(3)}}}_{\psi^2}{}^{ab\,i}{\!}_j$ contain terms quadratic and cubic in the bosonic fields, respectively. The higher-order terms, denoted by the dots, are fermionic. The expressions of the quadratic parts are
\begin{align}\label{N-quad}
\LL^{\text{\tiny{(2)}}}_{\psi^2}{}^{ab}{\!}_{ij} = &\,\tfrac12 \varepsilon_{iklm}\ga^{ab}E^{nm}D_{jn}{\!}^{kl}-\tfrac{1}{2}\varepsilon_{ijkl}\ga^{[a}\ga_{cd}\ga^{b]}R(V)^{cd}{}^{l}{\!}_{m}E^{mk}-2\ga^{ab}{P}^{c}D_{c}E_{ij}-\ga^{ab}D^{c}{P}_{c}E_{ij} \nonumber \\[1mm]
& \,-2\,T^{ab}{\!}_{kl}D_{ij}{\!}^{kl}+\tfrac12 \gamma^{[a}\ga_{ef}\ga_{cd}\gamma^{b]}{R}(V)^{ef}{}^{k}{\!}_{j}T^{cd}{\!}_{ik}-\tfrac12 \gamma^{[a}\ga_{cd}\ga_{ef}\gamma^{b]}{R}(V)^{ef}{}^{k}{\!}_{i}T^{cd}{\!}_{kj}\nonumber \\[1mm]
& \,+4\,{R}(M)^{abcd}T_{cd}{}_{ij}+4\,\varepsilon_{ijkl}\varepsilon^{abcd}{P}_{c}D^{e}T_{ed}{\!}^{kl}-4\,\varepsilon_{ijkl}{P}^{c}D_{c}T^{ba}{}^{kl}\nonumber\\
&\,-2\,\varepsilon_{ijkl}D_c {P}^{c}T^{ba}{}^{kl}\,,\\
\LL^{\text{\tiny{(2)}}}_{\psi^2}{}^{ab\,i}{\!}_j=&\,-4\,\varepsilon^{abcd}\delta^{i}{\!}_{j}\gamma_{d}D^{e}\left[{P}_{[c}\bar{P}_{e]}\right]\,.
\end{align}
while the cubic parts read
\begin{align}\label{N-cub}
\LL^{\text{\tiny{(3)}}}_{\psi^2}{}^{ab}{\!}_{ij}  =&\,\tfrac13 T^{ab}{\!}_{l[i}E_{j]k}E^{kl}-\tfrac23 T^{ab}{\!}_{ij}E_{kl}E^{kl}+\tfrac12 \varepsilon_{ijkl}\varepsilon_{mnpq}E^{ml}E^{pk}T^{ab}{}^{qn}+8\,T^{ba}{}^{kl}T_{lj}\cdot T_{ik} \nonumber \\[1mm]
&\, +16\,\ga\cdot T^{kl}T_{lj}{\!}^{[a}{\!}_{c}T^{b]c}{\!}_{ik}+ 8\,\bar{P}_{c}{P}^{[a}T^{b]c}{\!}_{ij}-16\,{P}_{c}\bar{P}^{[a}T^{b]c}{\!}_{ij}+4\,\varepsilon^{klmn}E_{n[i}T_{j]m}{}_{c}{\!}^{[a}T^{b]c}{\!}_{kl}\nonumber \\[1mm]
& \,+\tfrac{1}{6}\varepsilon_{klm(i}T_{j)n}{\!}^{ba}\ga\cdot T^{lm}E^{kn}\,,\\
\LL^{\text{\tiny{(3)}}}_{\psi^2}{}^{ab\,i}{\!}_j=&\,-2\,\gamma_{c}E^{ki}\bar{P}^{[a}T^{b]c}{\!}_{kj}-\gamma^{c}E^{ki}\bar{P}_{c}T^{ab}{\!}_{kj}- 2\,\gamma_{c}E_{kj}{P}^{[a}T^{b]c}{}^{ki}- \gamma^{c}E_{kj}{P}_{c}T^{ab}{}^{ki}\nonumber \\[1mm]
&\, +\tfrac{16} {3}\varepsilon^{ilkm}\gamma^{c}T^{ab}{\!}_{jl}\bar{P}^{d}T_{cdkm}+8\,\varepsilon^{iklm}\gamma^{[b}T^{a]c}{\!}_{lm}\bar{P}^{d}T_{cd}{}_{jk}-4\,\varepsilon^{iklm}\gamma^{c}\bar{P}^{[b}T^{a]d}{\!}_{lm}T_{dc}{}_{jk}\nonumber \\[1mm]
&\,-2\,\varepsilon^{iklm}\gamma^{[a}\bar{P}^{b]}T_{jk}.T_{lm}-\tfrac{16}{3}\varepsilon_{jklm}\gamma_{c}T^{ab}{}^{ik}{P}_{d}T^{dc}{}^{lm}+8\,\varepsilon_{jklm}\gamma^{[b}T^{a]c}{}^{lm}{P}^{d}T_{cd}{\!}^{ik} \nonumber \\[1mm]
&\, -4\,\varepsilon_{jklm}\gamma^{c}{P}^{[b}T^{a]d}{}^{lm}T_{dc}{\!}^{ik}-2\,\varepsilon_{jklm}\gamma^{[a}{P}^{b]}T^{ik}\cdot T^{lm}\,.
\end{align}

Finally we present the results for $\LL_{\psi\phi}{\!}^{ab}{\!}_{ij}$ and $\LL_{\psi\phi}{\!}^{ab\,i}{\!}_j$ which are coupled to a gravitino and an S-gauge field. The only bosonic terms are clearly at most quadratic
\begin{align}\label{NN-quad}
\LL_{\psi\phi}{\!}^{ab}{\!}_{ij}=&\,2\,\varepsilon_{ijkl}\varepsilon^{abcd}\gamma_{d}{P}^{e}T_{ec}{\!}^{kl}\,,\\
\LL_{\psi\phi}{\!}^{ab}{\!}_{ij}=&\,-4\,\delta^i{}_{j}{P}^{[a}\bar{P}^{b]}\,.
\end{align}

\setcounter{equation}{0}

\section{Transformations of the superconformal curvatures and Bianchi identities}
\label{App:trans-cur}

As mentioned in section \ref{sec:Preliminaries}, the gauge fields $\omega_\mu{}^{ab},f_\mu{}^a$ and $\phi_\mu{}^a$ are composite. They are expressed in terms of the other fields through the set of constraints \eqref{eq:cons-curv1}. The latter, when combined with the superconformal Bianchi identities, lead to the following useful relations
\begin{align}
\label{eq:Bian-id}
R(D)_{ab}=\,&0\,,\nonumber\\
R(M)_{abcd}=\,&R(M)_{cdab}\,,\nonumber\\
\varepsilon^{aecd}R(M)_{cdeb}=\,&0\,,\nonumber\\
\tfrac14\varepsilon^{abcd}\varepsilon^{efgh}R(M)_{cdgh}=\,&R(M)^{abef}\,,\nonumber\\
\varepsilon^{cdef}D_bD_dR(M)_{efab}=\,&0\,,\nonumber\\
R(K)_{ab}{}^c=\,&D_e R(M)_{ab}{}^{ec}\,,\nonumber\\
\varepsilon^{abcd}D_b R(V)_{cd}{}^i{}_j=\,&-\tfrac14\varepsilon^{iklm}\bar\Lambda_m\ga_b\ga\cdot T_{jl}R(Q)^{ab\,k}-(\text{h.c.; traceless})\,,\nonumber\\
D_a R(Q)^{ab\,i}=\,&-\tfrac14\varepsilon^{abcd}\ga_a R(S)_{cd}{}^i\nonumber\\
R(Q)^{+}_{ab}{}^{i}=\,&0\,,\nonumber\\
R(S)^{-}_{ab}{}^{i}=\,&\Dslash R(Q)_{ab}{}^i\,,\nonumber\\
\ga^{ab}R(S)_{ab}{}^i=\,&0\,,\nonumber\\
\ga^a R(S)^{+}_{ab}{}^{i}=\,&0\,,\nonumber\\
\varepsilon^{abcd}D_{b}R(S)_{cd}{}^{i}=\,& -\tfrac13\gamma^{a}T^{ij}\cdot R(S)_{j}-\tfrac43T^{ab}{}^{ij}D^{d}R(Q)_{db\,j}-\tfrac13\gamma^{a}R(V)^{i}
{}_{j}.R(Q)^{j} \nonumber \\
\,& -\tfrac16\mathrm{i}\gamma^{a}F\cdot R(Q)^{i}+\tfrac43 D^{g}T_{gc}{}^{ij}R(Q)^{ac}{}_{j}-\tfrac14\ga\cdot T_{jk}\gamma^{a}T^{ij}.R(Q)^{k}\,.
\end{align}
Note however that these relations are not independent. We recall that the (anti-)self dual part of a curvature is defined here as $R_{ab}^{\pm}=\,\tfrac12(R_{ab}\pm \tfrac12 \varepsilon_{abcd}R^{cd})$.  

The Q-supersymmetry and S-supersymmetry transformations of the supercovariant curvatures are
\begin{align}
\label{eq:trans-curv}
\delta_Q R(M)_{abcd}=\,&-\tfrac{1}{4}\eb^i\ga_{ab}R(S)^{-}_{cdi}-\tfrac{1}{4}\eb^i\ga_{cd}R(S)^{-}_{abi}+\tfrac{1}{4}\eb^i\Dslash\ga_{ab}R(Q)_{cdi}+\tfrac{1}{4}\eb^i\Dslash\ga_{cd}R(Q)_{abi}+\text{h.c.}\,,\nonumber\\
\delta_Q R(Q)_{ab}{}^i=\,&-\tfrac12 R(M)_{abcd}\ga^{cd}\epsilon^i+\tfrac14\big[\ga^{cd}\ga_{ab}+\tfrac13\ga_{ab}\ga^{cd}\big]\big[R(V)_{cd}{}^i{\!}_j\epsilon^j+\tfrac{1}{2} \mathrm{i}F_{cd}\epsilon^i+\Dslash T_{cd}{\!}^{ij}\epsilon_j\big]\,,\nonumber\\
\delta_Q R(V)_{ab}{}^i{}_j=\,&\eb^i R(S)_{abj}-2\eb^k\ga_{[a}D_{b]}\chi^i{\!}_{kj}+2\eb_l\chi^i{\!}_{kj}T_{ab}{\!}^{kl}+\tfrac13 T_{abjl}\big[-2\,\eb^l \slashed{\bar{P}}\Lambda^i-\eb^lE^{ik}\Lambda_k\big]\nonumber\\
\,&+\tfrac{1}{8}\varepsilon^{iklm}\eb^n\ga_{[a}\ga \cdot T_{kn}\ga \cdot T_{lj}\ga_{b]}\Lambda_m-\tfrac12E^{ik}\varepsilon_{jkmn}\eb^mR(Q)_{ab}{\!}^n\nonumber\\
\,&-\tfrac14\varepsilon^{iklp}\varepsilon_{jmnp}\eb^m\ga^cR(Q)_{abk}\bar\Lambda_l\ga_c\Lambda^n+\tfrac13\eb_j\ga_{[a}D_{b]}\big[E^{ik}\Lambda_k\big]\nonumber\\
\,&+\tfrac12\varepsilon^{iklm}\eb_{k}D_{[a}\big[\ga\cdot T_{lj}\ga_{b]}\Lambda_m\big]+\tfrac23\eb_j\ga_{[a}D_{b]}\big[\slashed{\bar{P}} \Lambda^i\big]-(\text{h.c.; traceless})\,,\nonumber\\
\delta_Q R(S)^{+}_{ab}{\!}^{i}=\,&-2D^c R(M)^{+}{\!}_{abcd}ga^d\epsilon^i+\tfrac14\big[\ga^{cd}\ga_{ab}+\tfrac13\ga_{ab}\ga^{cd}\big]\big[\ga^e\epsilon^jD_e R_{cd}{}^i{\!}_j+\tfrac{1}{2}\mathrm{i}\ga^e\epsilon^jD_e F_{cd}\nonumber\\
\,&+\epsilon_j D_c D^e T_{ed}{\!}^{ij}+2\epsilon_l T_{cdjk}T^{ij}.T^{kl}-4\ga^e\epsilon^k D^fT_{fe}{\!}^{ij}T_{cdjk}-2\ga^e\epsilon^kD^{f}T_{cdjk}T_{fe}{\!}^{ij}\big]\,,\nonumber\\
\delta_S R(M)_{abcd}=\,&-\tfrac{3}{4}\bar\eta_i \ga_{ab}R(Q)_{cd}{\!}^i-\tfrac{3}{4}\bar\eta_i \ga_{cd}R(Q)_{ab}{\!}^i+\text{h.c.}\,,\nonumber\\
\delta_S R(Q)_{ab}{\!}^i=\,&\tfrac12\big[\ga^{cd}\ga_{ab}+\tfrac13\ga_{ab}\ga^{cd}\big]T_{cd}{\!}^{ij}\eta_j\,,\nonumber\\
\delta_S R(V)_{ab}{}^i{\!}_j=\,&\bar\eta^iR(Q)_{abj}+\varepsilon^{iklm}T_{abjl}\bar\eta_{k}\Lambda_m-\bar\eta^k\ga_{ab}\chi^i{\!}_{kj}-\tfrac16\bar\eta^i\ga_{ab}\big[2\Pslash\Lambda_{j}-E_{jk}\Lambda^k\big]\nonumber\\
\,&-(\text{h.c.; traceless})\,,\nonumber\\
\delta_S R(S)^{+}_{ab}{\!}^{i}=\,&-\tfrac12 R(M)_{abcd}\ga^{cd}\eta^i+\tfrac{3}{4}\big[\ga^{cd}\ga_{ab}+\tfrac13\ga_{ab}\ga^{cd}\big]\big[\eta^j R(V)_{cd}{}^i{\!}_j+\tfrac{1}{2}\mathrm{i}\eta^i F_{cd}\big]\,.
\end{align}
The transformations of $R_{ab}{}^c(K)$ and $R_{ab}{}^{i(-)}(S)$ can be easily derived from \eqref{eq:Bian-id}. Finally, for the purpose of section \ref{sec:strategy}, we give the explicit expressions of the fermionic supercovariant curvatures
\begin{align}
R(Q)_{\mu\nu}{\!}^{i}=&\,2\,\mathcal{D}_{[\mu}\psi_{\nu]}{\!}^{i}-\gamma_{[\mu}\phi_{\nu]}{\!}^{i}-\tfrac{1}{2}\gamma\cdot T^{ij}\gamma_{[\mu}\psi_{\nu]j}+\tfrac{1}{2}\varepsilon^{ijkl}\bar{\psi}_{\mu j}\psi_{\nu k}\Lambda_{l} \label{eq:Qcurv} \\[1mm]
R(S)_{\mu\nu}{\!}^{i}=&\,2\,\mathcal{D}_{[\mu}\phi_{\nu]}{\!}^{i}-2f_{[\mu}{\!}^{a}\gamma_{a}\psi_{\nu]}{\!}^{i}-\tfrac{1}{6}\gamma_{[\mu}\gamma \cdot T^{ij}\phi_{\nu]j}-\tfrac{1}{2}\varepsilon^{ijkl}\bar{\phi}_{[\mu k}\Lambda_{l}\psi_{\nu]j}+\delta^{\scriptscriptstyle{(cov)}}_{Q\vert_{ \psi_{[\nu}}}\phi_{\mu]}{\!}^{i}\nonumber\\
&\,+[\text{terms}\propto\psi^2] \label{eq:Scurv}
\end{align} 
where the symbol $\delta^{\scriptscriptstyle{(cov)}}_{Q\vert_{\psi_{b}}}$ denotes the supercovariant part of a Q-variation with the parameter replaced by the gravitino.

\end{appendix}

\providecommand{\href}[2]{#2}

\end{document}